\documentclass[preprint,prd,aps,showpacs,showkeys,nofootinbib]{revtex4}
\usepackage{graphicx}
\begin{document}

\title{B meson rare decays in the TNMSSM}

\author{Hai-Xiang Chen$^{a}$\footnote{haixchen@hotmail.com},
Sheng-Kai Cui$^{a}$,
Ning-Yu Zhu$^{a}$,
Zhao-Yang Zhang$^{a}$,
Huai-Cong Hu$^{a}$}

\affiliation{$^a$Department of Physics, Guangxi University, Nanning, 530004, China}

\begin{abstract}
We investigate the two loop electroweak corrections to B meson rare decays $\bar B\rightarrow X_s\gamma$ and $B_s^0\rightarrow \mu^+\mu^-$ in the minimal supersymmetry standard model (MSSM) extension with two triplets and one singlet (TNMSSM). The new particle contents and interactions in the TNMSSM can affect the theoretical predictions of the branching ratios ${\rm Br}(\bar B\rightarrow X_s\gamma)$ and ${\rm Br}(B_s^0\rightarrow \mu^+\mu^-)$, and the corrections from two loop diagrams to the process $\bar B\rightarrow X_s\gamma$ can reach around $4\%$. Considering the latest experimental measurements, the numerical results of ${\rm Br}(\bar B\rightarrow X_s\gamma)$ and ${\rm Br}(B_s^0\rightarrow \mu^+\mu^-)$ in the TNMSSM are presented and analyzed. It is found that the results in the TNMSSM can fit the updated experimental data well and the new parameters $T_{\lambda},\;\kappa,\;\lambda$ affect the theoretical predictions of ${\rm Br}(\bar B\rightarrow X_s\gamma)$ and ${\rm Br}(B_s^0\rightarrow \mu^+\mu^-)$ obviously.

\end{abstract}


\maketitle

\section{Introduction\label{sec1}}
\indent\indent
After the discovery of Higgs on the Large Hadron Collider (LHC) in 2012, all particles predicted by the SM have been found. However, there are still some problems that are difficult to be solved by the SM, such as the non-zero neutrino mass, resonable dark matter candidates and etc. It indicates new physics is needed to extend the SM. And since B meson rare decay processes $\bar B\rightarrow X_s\gamma$ and $B_s^0\rightarrow \mu^+\mu^-$ are not affected by the uncertainties of non-perturbative QCD, research on B physics is particularly sensitive to exploring new physical effects beyond the SM. In Refs.~\cite{17,BaBar1,BaBar2,Belle}, the average experimental data on the branching ratios of $\bar B\rightarrow X_s\gamma$ and $B_s^0\rightarrow \mu^+\mu^-$ are provided as
\begin{eqnarray}
&&{\rm Br}(\bar B\rightarrow X_s \gamma)=(3.49\pm0.19)\times 10^{-4},\nonumber\\
&&{\rm Br}(B_s^0\rightarrow \mu^+\mu^-)=(2.9_{-0.6}^{+0.7})\times10^{-9}.
\label{experimental data}
\end{eqnarray}
The branching ratios of $\bar B\rightarrow X_s\gamma$ and $B_s^0\rightarrow \mu^+\mu^-$ predicted by the SM are~\cite{SMP1,SMP2,SMP3,SMP4,SMP5,SMP6,SMP7,SMP8,SMP9}
\begin{eqnarray}
&&{\rm Br}(\bar B\rightarrow X_s \gamma)_{\rm SM}=(3.36\pm0.23)\times 10^{-4},\nonumber\\
&&{\rm Br}(B_s^0\rightarrow \mu^+\mu^-)_{\rm SM}=(3.23\pm0.27)\times10^{-9},
\end{eqnarray}
which coincides with the experimental data very well. Therefore, the new physics contributions to $\bar B\rightarrow X_s\gamma$ and $B_s^0\rightarrow \mu^+\mu^-$ are limited strictly by the accurate measurements on the processes $\bar B\rightarrow X_s\gamma$ and $B_s^0\rightarrow \mu^+\mu^-$.

As one of the most famous extensions of the SM, the B meson rare decay process $\bar B\rightarrow X_s\gamma$ is analyzed in the MSSM~\cite{B4,B5,B6,B7,B8,B9,B10,B11}. In 1998, Ciuchini presented the QCD corrections to $\bar B\rightarrow X_s\gamma$ at Next-to-leading order (NLO) in the Two-Higgs doublet model (THDM)~\cite{NPB1}. Then, the two loop QCD corrections was proposed in Ref.~\cite{NPB2}. Not only the process $\bar B\rightarrow X_s\gamma$, there are also many references researching on other B meson rare decay processes in the THDM~\cite{He:1988tf,Skiba:1992mg,Choudhury:1998ze,Huang:2000sm,Crivellin2013,Crivellin2019}. Recently, the authors of Refs.~\cite{NPB3,NPB4,NPB5,NPB6,NPB7,Zhang1,Feng1,Feng2,Feng3,Yang1,Chen1} have discussed the supersymmetric effects on the B meson rare decay processes. Meanwhile, Long et al present the computation of the flavor transition process $b\rightarrow s\gamma$~\cite{Long1}. The authors of Ref.~\cite{NPB11} have investigated the two aspects of hadronic B decays, and then these processes in the case of CP violation have been discussed~\cite{NPB12}. Moreover, many possibilities for searching supersymmetry effects in different B meson rare decay processes have been proposed~\cite{NPB13,NPB14,NPB15}. When investigating the processes of rare B decay, the supersymmetry effects are very interesting, and B decay can be conducive for us to understand the characteristics of the supersymmetry model in detail while limiting the parameter space~\cite{NPB16,NPB17}.

TNMSSM is an extension of Next-to minimal supersymmetric standard model (NMSSM) containing two $SU(2)_L$ triplets with hypercharge $\pm1$, where the NMSSM introduces an additional scalar singlet compared with the MSSM~\cite{5}. The new scalar singlet in the NMSSM is introduced to solve the $\mu$ problem in the MSSM~\cite{NMSSM1,NMSSM2}. However, NMSSM fails to improve the little hierarchy problem~\cite{NMSSM3,NMSSM4,NMSSM5,NMSSM6,NMSSM7,NMSSM8}. Fortunately, the author of Ref.~\cite{5} solved these problems in the TNMSSM by introducing two scalar triplets which are responsible for large correction to the lightest physical Higgs mass. The tiny neutrino mass measured at the neutrino oscillation experiments can be obtained by applying type \uppercase\expandafter{\romannumeral2} seesaw mechanism and a discrete flavor symmetry $G_F$ in TNMSSM (i.e. the flavored-TNMSSM)~\cite{flaTNMSSM}. In this work, we analyze the two loop electroweak corrections to $\bar B\rightarrow X_s\gamma$ and $B_s^0\rightarrow \mu^+\mu^-$ in the TNMSSM. Compared with the MSSM, new particle contents and interactions can make important contributions to the processes.

The paper is organized as follows. The superpotential, the soft breaking terms and the mass matrices of singly-charged Higgs and CP-even Higgs in the TNMSSM are reviewed briefly in Sec.II. Sec.III and Sec.IV give the corresponding Wilson coefficients and analytic expressions for ${\rm Br}(\bar B\rightarrow X_s\gamma)$ and ${\rm Br}(B_s^0\rightarrow \mu^+\mu^-)$. Sec.V analyses the numerical results and Sec.VI gives a summary. The corresponding matrix elements and the concrete expressions of the Wilson coefficients are collected in the appendices.

\section{The TNMSSM\label{sec2}}
\indent\indent
There are different versions of TNMSSM, such as the type of a triplet with hypercharge $Y=0$~\cite{1Y0,2Y0,3Y0}. Here we adopt the version of two triplets with hypercharge $\pm1$ to keep on our work. As mentioned before, the TNMSSM contains two $SU(2)_L$ triplets $\hat T\sim(1,3,1)$, $\hat {\bar T}\sim(-1,3,1)$ with $Y=\pm1$ and a SM gauge singlet $\hat S\sim(0,1,1)$. For the mass matrices and the interaction vertexes needed, we encode this version of TNMSSM in SARAH~\cite{SAR1,SAR2,SAR3,SAR4,SAR5}. The chiral superfields for quarks and leptons are given by
\begin{eqnarray}
&&\hat{Q}=\left(\begin{array}{c}\hat U\\ \hat D\end{array}\right)\sim(1/6, 2, 3),\quad\; \hat{L}=\left(\begin{array}{c}\hat {\nu}\\ \hat E\end{array}\right)\sim(-1/2, 2, 1), \nonumber\\
&&\hat{U}^c\sim(-2/3, 1, 3),\quad\; \hat{D}^c\sim(1/3, 1, 3),\quad\; \hat E^c\sim(1, 1, 1),
\end{eqnarray}
where we ignore the index of generations and the quantum numbers of $U(1)_Y$, $SU(2)_L$ and $SU(3)_C$ are indicated in the bracket, respectively. Additionally, the expressions and the quantum numbers of two Higgs triplets, two doublets and one singlet are assigned as
\begin{eqnarray}
&&\hat{T}=\left(\begin{array}{cc}\frac{1}{\sqrt2}T^+,&-T^{++}\\ T^0,&\frac{-1}{\sqrt2}T^+\end{array}\right)\sim (1, 3, 1),\quad\; \hat{\bar T}=\left(\begin{array}{cc}\frac{1}{\sqrt2}{\bar T}^-,&-{\bar T}^0\\ {\bar T}^{--},&\frac{-1}{\sqrt2}{\bar T}^-\end{array}\right)\sim (-1, 3, 1), \nonumber\\
&&\hat{H_d}=\left(\begin{array}{c}H_d^0\\ H_d^-\end{array}\right)\sim (-1/2, 2, 1),\quad\;\hat{H_u}=\left(\begin{array}{c}H_u^+\\ H_u^0\end{array}\right)\sim(1/2, 2, 1),\quad\;\hat S \sim(0, 1, 1).
\end{eqnarray}
In the previous expressions, $T^0$ and ${\bar T}^0$ are two complex neutral superfields, while $T^+$, $\bar T^-$ are singly-charged Higgs and $T^{++}$, $\bar T^{--}$ are doubly-charged Higgs.

The superpotential of the TNMSSM $W_{\rm TNMSSM}$ contains two parts
\begin{eqnarray}
&&W_{\rm TNMSSM}=W_{\rm MSSM}+W_{\rm TS},
\end{eqnarray}
with
\begin{eqnarray}
&&W_{\rm MSSM}=Y_u \hat U^c \hat H_u \cdot \hat Q -Y_d \hat D^c \hat H_d \cdot \hat Q -Y_e \hat E^c \hat H_d \cdot \hat L,
\label{WMSSM}
\end{eqnarray}
where $W_{\rm MSSM}$ is the superpotential of the MSSM, and $W_{\rm TS}$ explains the extended scalar sector including two triplets and a SM gauge singlet,
\begin{eqnarray}
&&W_{\rm TS}=\chi_d \hat H_d \cdot \hat T \hat H_d +\chi_u \hat H_u \cdot \hat {\bar T} \hat H_u +\frac{1}{3}\kappa \hat S \hat S \hat S +\lambda \hat S \hat H_u \cdot \hat H_d +\Lambda_T \hat S \mathrm{Tr}(\hat {\bar T} \hat T).
\end{eqnarray}
Here, we also neglect the index of generations. From Eq.(\ref{WMSSM}), we could find that there are only two MSSM Higgs doublets coupled with fermion multiplet via Yukawa coupling. Then, the general soft breaking terms are given by
\begin{eqnarray}
&&\mathcal{-L}_{\rm soft}=m_{H_u}^2 |H_u|^2 +m_{H_d}^2 |H_d|^2 +m_S^2 |S|^2 +m_{T}^2 \mathrm{Tr}(|T|^2) +m_{\bar T}^2 \mathrm{Tr}(|\bar T|^2)
\nonumber\\
&&\hspace{1.4cm}
+m_{Q}^2 |Q|^2 +m_{U}^2 |U|^2 +m_{D}^2 |D|^2 +(T_{\Lambda_T} S \mathrm{Tr}(T \bar T) +T_{\lambda} S H_u \cdot H_d +\frac{1}{3} T_{\kappa} S^3
\nonumber\\
&&\hspace{1.4cm}
-T_{\chi_u} H_u \cdot \bar T H_u -T_{\chi_d} H_d \cdot T H_d + T_{u,ij} \tilde{Q_j} \cdot H_u \tilde{U^c_i} -T_{d,ij} \tilde{Q_j} \cdot H_d \tilde{D^c_i} +H.c.),
\end{eqnarray}
where
\begin{eqnarray}
&&H_u \cdot H_d =H_u^+ H_d^- -H_u^0 H_d^0, \\
&&H_d \cdot T H_d=\sqrt{2} H_d^- H_d^0 T^+ -(H_d^0)^2 T^0 -(H_d^-)^2 T^{++}, \\
&&H_u \cdot \bar T H_u=\sqrt{2} H_u^+ H_u^0 {\bar T}^- -(H_u^0)^2 {\bar T}^0 -(H_u^+)^2 {\bar T}^{--}.
\end{eqnarray}
When the $Z_3$ symmetry is imposed, the $\mu$ term only forms after the singlet $S$ has obtained a vacuum expectation value (VEV) $v_S$ as $\mu=\frac{1}{\sqrt2} \lambda v_S$. The coefficients in the Higgs sector are assumed to be real in the following calculations.

The $SU(2)_L\bigotimes U(1)_Y$ electroweak symmetry breaking occurs when the neutral parts of Higgs fields obtain the VEVs
\begin{eqnarray}
&&H_d^0=\frac{1}{\sqrt2} (v_d + \Re {H_d^0} +i \Im {H_d^0}),\quad\; H_u^0=\frac{1}{\sqrt2} (v_u +\Re {H_u^0} +i \Im {H_u^0}), \nonumber\\
&&T^0=\frac{1}{\sqrt2} (v_T + \Re {T^0} +i \Im {T^0}),\quad\; {\bar T}^0=\frac{1}{\sqrt2} (v_{\bar T} +\Re {{\bar T}^0} +i \Im {{\bar T}^0}), \nonumber\\
&&S=\frac{1}{\sqrt2} (v_S +\Re S +i \Im S).
\end{eqnarray}
Meanwhile, the VEVs of the triplets must to be small to avoid large $\rho$ parameter correction~\cite{5} and the VEV of the singlet is required to be large for generating a large $\mu$ term like the case in the NMSSM~\cite{NMSSM1}.
In the TNMSSM, the mass of $Z$ gauge boson reads
\begin{eqnarray}
&&M_Z^2=\frac{1}{4} (g_1^2 +g_2^2) (v_u^2+v_d^2 +4 v_T^2 +4 v_{\bar T}^2)=\frac{1}{4} (g_1^2 +g_2^2) v^2,
\end{eqnarray}
where $g_1$ and $g_2$ represent the gauge coupling constants of $U(1)_Y$ and $SU(2)_L$ respectively. It will be seen from this that due to the triplets, the electroweak symmetry breaking VEV for the doublets compared to MSSM is changed as
\begin{eqnarray}
&&v=\sqrt {v_u^2+v_d^2 +4 v_T^2 +4 v_{\bar T}^2} \approx 246\; \rm GeV.
\end{eqnarray}

Minimizing the Higgs scalar potential
\begin{eqnarray}
&&\frac{\partial V}{\partial {v_u}}=\frac{\partial V}{\partial {v_d}}=\frac{\partial V}{\partial {v_T}}=\frac{\partial V}{\partial {v_{\bar T}}}=\frac{\partial V}{\partial {v_S}}=0,
\end{eqnarray}
we can deduce the squared mass matrices of the neutral Higgs and singly-charged Higgs. In the basis $(H_d^{-}, H_u^{+,\ast}, {\bar T}^{-}, T^{+,\ast})$ and $(H_d^{-,\ast}, H_u^{+}, {\bar T}^{-,\ast}, T^{+})$, the squared mass matrix for singly-charged Higgs can be expressed as
\begin{eqnarray}
&&M_{H^{\pm}}^2=\left(\begin{array}{*{20}{c}}
m_{H_d^- H_d^{-,\ast}}&m^{\ast}_{H_u^{+,\ast} H_d^{-,\ast}}&m^{\ast}_{{\bar T}^- H_d^{-,\ast}}&m^{\ast}_{T^{+,\ast} H_d^{-,\ast}}\\ [2pt]
m_{H_d^- H_u^+}&m_{H_u^{+,\ast} H_u^+}&m^{\ast}_{{\bar T}^- H_u^+}&m^{\ast}_{T^{+,\ast} H_u^+}\\ [2pt]
m_{H_d^- {\bar T}^{-,\ast}}&m_{H_u^{+,\ast} {\bar T}^{-,\ast}}&m_{{\bar T}^- {\bar T}^{-,\ast}}&m^{\ast}_{T^{+,\ast} {\bar T}^{-,\ast}}\\ [2pt]
m_{H_d^- T^+}&m_{H_u^{+,\ast} T^+}&m_{{\bar T}^- T^+}&m_{T^{+,\ast} T^+}
\end{array}\right).
\label{MCH}
\end{eqnarray}

The $4 \times 4$ squared mass matrix in Eq.(\ref{MCH}) can be diagonalized by the unitary matrix $Z^{H^{\pm}}$
\begin{eqnarray}
&&Z^{H^{\pm}} M_{H^{\pm}}^2 Z^{H^{\pm},\dagger}=M_{H^{\pm},dia}^2.
\end{eqnarray}
Then, we can get three mass eigenstates $(H_1^{\pm},H_2^{\pm},H_3^{\pm})$ for singly-charged Higgs and one state $G^{\pm}$ for the massless Goldstone boson. Therefore, the TNMSSM has two more singly-charged Higgs due to the triplets, with respect to MSSM. These new defined singly-charged Higgs will bring contributions to loop corrections for $\bar B\rightarrow X_s\gamma$ and $B_s^0\rightarrow \mu^+\mu^-$.

At tree level, the squared mass matrix for CP-even Higgs is given in the basis ($\Re H_d^0$, $\Re H_u^0$, $\Re S$, $\Re T^0$, $\Re {\bar T}^0$) as
\begin{eqnarray}
&&M_h^2=\left(\begin{array}{*{20}{c}}
m_{H_d^0 H_d^0}&m_{H_u^0 H_d^0}&m_{S H_d^0}&m_{T^0 H_d^0}&m_{{\bar T}^0 H_d^0}\\ [6pt]
m_{H_d^0 H_u^0}&m_{H_u^0 H_u^0}&m_{S H_u^0}&m_{T^0 H_u^0}&m_{{\bar T}^0 H_u^0}\\ [6pt]
m_{H_d^0 S}&m_{H_u^0 S}&m_{S S}&m_{T^0 S}&m_{{\bar T}^0 S}\\ [6pt]
m_{H_d^0 T^0}&m_{H_u^0 T^0}&m_{S T^0}&m_{T^0 T^0}&m_{{\bar T}^0 T^0}\\ [6pt]
m_{H_d^0 {\bar T}^0}&m_{H_u^0 {\bar T}^0}&m_{S {\bar T}^0}&m_{T^0 {\bar T}^0}&m_{{\bar T}^0 {\bar T}^0}
\end{array}\right).
\end{eqnarray}
The corresponding matrix elements of the squared mass matrices for singly-charged Higgs and CP-even Higgs are collected in Appendix A.

Including the leading-log radiative corrections up to two loops for stop and top sector~\cite{HiggsC1,HiggsC2,HiggsC3}, the mass of the SM-like Higgs boson can be written as
\begin{eqnarray}
&&m_h=\sqrt{(m_{h_1}^0)^2+\Delta m_h^2},\label{higgs mass} \nonumber\\
&&\Delta m_h^2=\frac{3m_t^4}{2\pi v^2}\Big[\Big(\tilde{t}+\frac{1}{2}+\tilde{X}_t\Big)+\frac{1}{16\pi^2}\Big(\frac{3m_t^2}{2v^2}-32\pi\alpha_3\Big)\Big(\tilde{t}^2
+\tilde{X}_t \tilde{t}\Big)\Big],\nonumber\\
&&\tilde{t}=log\frac{M_S^2}{m_t^2},\qquad\;\tilde{X}_t=\frac{2\tilde{A}_t^2}{M_S^2}\Big(1-\frac{\tilde{A}_t^2}{12M_S^2}\Big),\label{higgs corrections}
\end{eqnarray}
where $m_{h_1}^0$ is the lightest tree-level Higgs boson mass, $\alpha_3$ is the running QCD coupling constant, $M_S=\sqrt{m_{\tilde t_1}m_{\tilde t_2}}$ is the geometric mean of the stop masses $m_{\tilde t_{1,2}}$, $m_t$ is the top quark pole mass, and $\tilde{A}_t=A_t-\mu \cot\beta$ with $A_t=T_{u,33}$ being the trilinear Higgs stop coupling.

In the basis $(\tilde u_L, \tilde u_R)$ and $(\tilde d_L, \tilde d_R)$, we can obtain the squared mass matrices for Up-Squark and Down-Squark as
\begin{eqnarray}
&&m_{\tilde u}^2=
\left(\begin{array}{cc}m_{\tilde u_L \tilde{u}_L^\ast}&m_{\tilde u_R \tilde{u}_L^\ast}^\dagger\\m_{\tilde u_L \tilde{u}_R^\ast}&m_{\tilde u_R \tilde{u}_R^\ast}\end{array}\right),
\end{eqnarray}
\begin{eqnarray}
&&m_{\tilde d}^2=
\left(\begin{array}{cc}m_{\tilde d_L \tilde{d}_L^\ast}&m_{\tilde d_R \tilde{d}_L^\ast}^\dagger\\m_{\tilde d_L \tilde{d}_R^\ast}&m_{\tilde d_R \tilde{d}_R^\ast}\end{array}\right),
\end{eqnarray}
where
\begin{eqnarray}
&&m_{\tilde u_L \tilde{u}_L^\ast}=-\frac{1}{24} (-3 g_2^2+g_1^2) (2 v_{\bar T}^2-2 v_T^2-v_u^2+v_d^2)+\frac{1}{2} (2 m_q^2+v_u^2 Y_u^\dagger Y_u), \nonumber\\
&&m_{\tilde u_L \tilde{u}_R^\ast}=\frac{1}{2} (\sqrt2 v_u T_u+Y_u (2 \chi_u v_u v_{\bar T}-\lambda v_d v_S)), \nonumber\\
&&m_{\tilde u_R \tilde{u}_R^\ast}=\frac{1}{6}g_1^2 (2 v_{\bar T}^2-2 v_T^2-v_u^2+v_d^2)+\frac{1}{2}(2 m_u^2+v_u^2 Y_u Y_u^\dagger), \nonumber\\
&&m_{\tilde d_L \tilde{d}_L^\ast}=-\frac{1}{24} (3 g_2^2+g_1^2) (2 v_{\bar T}^2-2 v_T^2-v_u^2+v_d^2)+\frac{1}{2} (2 m_q^2+v_d^2 Y_d^\dagger Y_d), \nonumber\\
&&m_{\tilde d_L \tilde{d}_R^\ast}=\frac{1}{2} (\sqrt2 v_d T_d+Y_d (2 \chi_d v_d v_T-\lambda v_u v_S)), \nonumber\\
&&m_{\tilde d_R \tilde{d}_R^\ast}=\frac{1}{12}g_1^2 (-2 v_{\bar T}^2+2 v_T^2+v_u^2-v_d^2)+\frac{1}{2}(2 m_d^2+v_d^2 Y_d Y_d^\dagger).
\end{eqnarray}


\section{Rare decay $\bar B\rightarrow X_s\gamma$ \label{sec3}}
\indent\indent
The effective Hamilton for the transition $b\rightarrow s$ at hadronic scale can be described by
\begin{eqnarray}
&&H_{eff}=-\frac{4G_F}{\sqrt{2}}V_{ts}^\ast V_{tb}\Big[C_1\mathcal{O}^c_1+C_2\mathcal{O}_2^c+\sum_{i=3}^6\mathcal{O}_i+\sum_{i=7}^{10}(C_i\mathcal{O}_i+C'_i\mathcal{O}'_i)\nonumber\\
&&\qquad\;\quad\;+\sum_{i=S,P}(C_i\mathcal{O}_i+C'_i\mathcal{O}'_i)\Big].
\end{eqnarray}
From the Refs.~\cite{O1,O2,Altmannshofer:2008dz,O3,O4,O6}, $\mathcal{O}_i(i=1,...,10,S,P)$ and $\mathcal{O}'_i(i=7,...,10,S,P)$ are defined as
\begin{eqnarray}
&&{\cal O}_{_1}^c=(\bar{s}_{_L}\gamma_\mu T^au_{_L})(\bar{u}_{_L}\gamma^\mu T^ab_{_L})\;,\;\;
{\cal O}_{_2}^c=(\bar{s}_{_L}\gamma_\mu u_{_L})(\bar{u}_{_L}\gamma^\mu b_{_L})\;,\nonumber\\
&&{\cal O}_{_3}=(\bar{s}_{_L}\gamma_\mu b_{_L})\sum\limits_q(\bar{q}\gamma^\mu q)\;,\;\;
{\cal O}_{_4}=(\bar{s}_{_L}\gamma_\mu T^ab_{_L})\sum\limits_q(\bar{q}\gamma^\mu T^aq)\;,\nonumber\\
&&{\cal O}_{_5}=(\bar{s}_{_L}\gamma_\mu\gamma_\nu\gamma_\rho b_{_L})\sum\limits_q(\bar{q}\gamma^\mu\gamma^\nu\gamma^\rho q)\;,\;\;
{\cal O}_{_6}=(\bar{s}_{_L}\gamma_\mu\gamma_\nu\gamma_\rho T^ab_{_L})\sum\limits_q(\bar{q}\gamma^\mu\gamma^\nu\gamma^\rho T^aq)\;,\nonumber\\
&&{\cal O}_{_7}={e\over 16\pi^2}m_{_b}(\bar{s}_{_L}\sigma_{_{\mu\nu}}b_{_R})F^{\mu\nu}\;,\;\;
{\cal O}_{_7}'={e\over 16\pi^2}m_{_b}(\bar{s}_{_R}\sigma_{_{\mu\nu}}b_{_L})F^{\mu\nu}\;,\nonumber\\
&&{\cal O}_{_8}={g_{_s}\over 16\pi^2}m_{_b}(\bar{s}_{_L}\sigma_{_{\mu\nu}}T^ab_{_R})G^{a,\mu\nu}\;,\;\;
{\cal O}_{_8}'={g_{_s}\over 16\pi^2}m_{_b}(\bar{s}_{_R}\sigma_{_{\mu\nu}}T^ab_{_L})G^{a,\mu\nu}\;,\nonumber\\
&&{\cal O}_{_9}={e^2\over g_{_s}^2}(\bar{s}_{_L}\gamma_\mu b_{_L})\bar{l}\gamma^\mu l\;,\;\;
{\cal O}_{_9}'={e^2\over g_{_s}^2}(\bar{s}_{_R}\gamma_\mu b_{_R})\bar{l}\gamma^\mu l\;,\nonumber\\
&&{\cal O}_{_{10}}={e^2\over g_{_s}^2}(\bar{s}_{_L}\gamma_\mu b_{_L})\bar{l}\gamma^\mu\gamma_5 l\;,\;\;
{\cal O}_{_{10}}'={e^2\over g_{_s}^2}(\bar{s}_{_R}\gamma_\mu b_{_R})\bar{l}\gamma^\mu\gamma_5 l\;,\nonumber\\
&&{\cal O}_{_S}={e^2\over16\pi^2}m_{_b}(\bar{s}_{_L}b_{_R})\bar{l}l\;,\;\;
{\cal O}_{_S}'={e^2\over16\pi^2}m_{_b}(\bar{s}_{_R}b_{_L})\bar{l}l\;,\;\;\nonumber\\
&&{\cal O}_{_P}={e^2\over16\pi^2}m_{_b}(\bar{s}_{_L}b_{_R})\bar{l}\gamma_5l\;,\;\;
{\cal O}_{_P}'={e^2\over16\pi^2}m_{_b}(\bar{s}_{_R}b_{_L})\bar{l}\gamma_5l\;,
\label{operators}
\end{eqnarray}
where $g_s$ represents the strong coupling, $F^{\mu\nu}$ is the electromagnetic field strength tensor, $G^{\mu\nu}$ is the gluon field strength tensor, and $T^a\,(a=1,...,8)$ are the $SU(3)$ generators.

\begin{figure}
\setlength{\unitlength}{1mm}
\centering
\includegraphics[width=6in,height=3in]{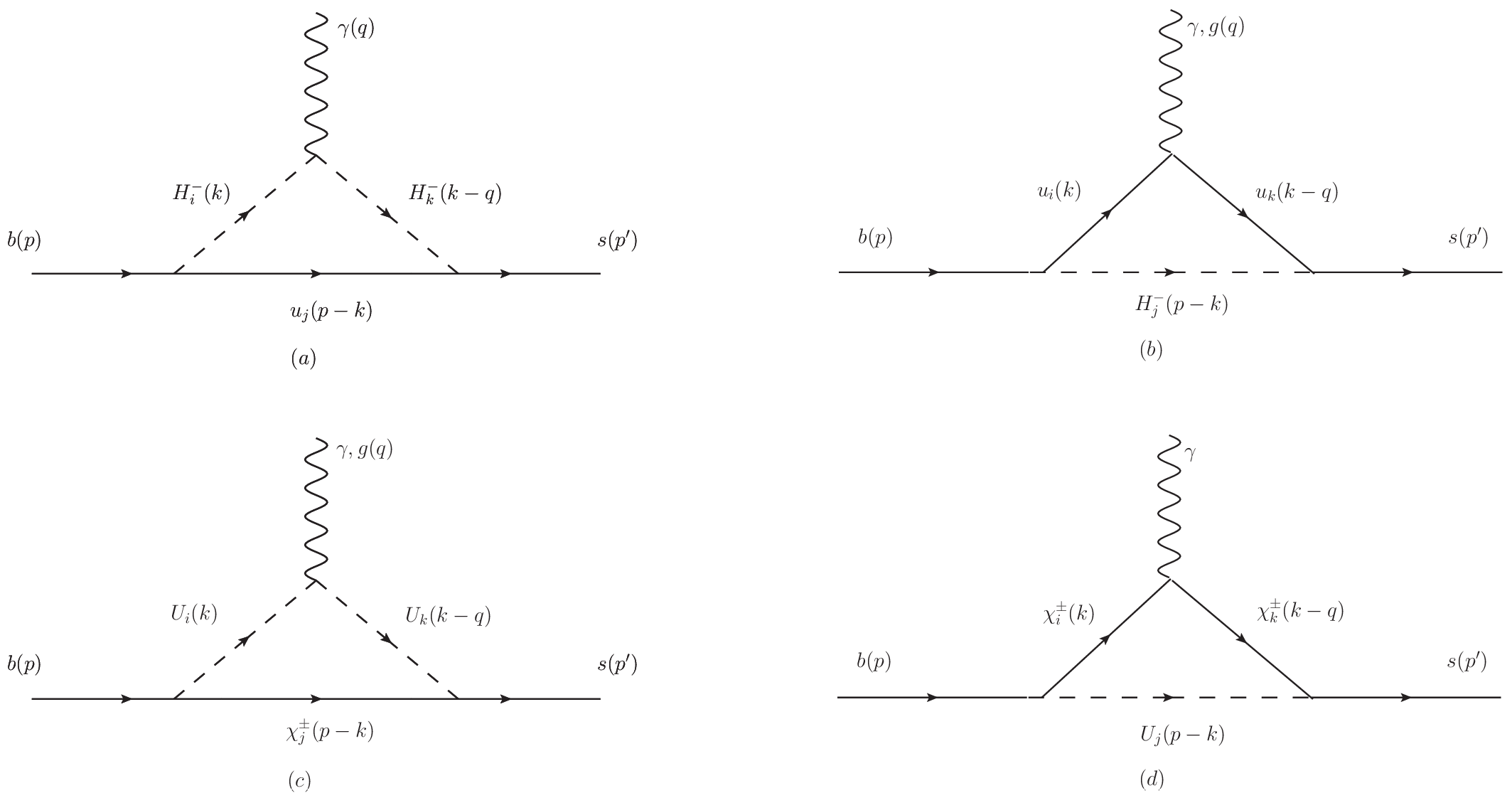}
\vspace{0cm}
\caption[]{The one loop Feynman diagrams contributing to $\bar{B}\rightarrow X_s\gamma$ in the TNMSSM.}
\label{figOne}
\end{figure}

As shown in Fig.~\ref{figOne}, the main one loop Feynman diagrams contributing to the process $\bar{B}\rightarrow X_s\gamma$ in the TNMSSM are mediated by newly defined up-squarks, singly-charged Higgs and charginos. Compared to the MSSM, these new definitions of particles will affect the prediction of the process $\bar{B}\rightarrow X_s\gamma$. In the TNMSSM, the one loop Wilson coefficients corresponding to Fig.~\ref{figOne} are $C_{7,NP}^{(a)}(\mu_{EW})$, $C_{7,NP}^{(b)}(\mu_{EW})$, $C_{7,NP}^{(c)}(\mu_{EW})$ and $C_{7,NP}^{(d)}(\mu_{EW})$ and their concrete expressions can be obtained from Appendix B.

\begin{figure}
\setlength{\unitlength}{1mm}
\centering
\includegraphics[width=6in]{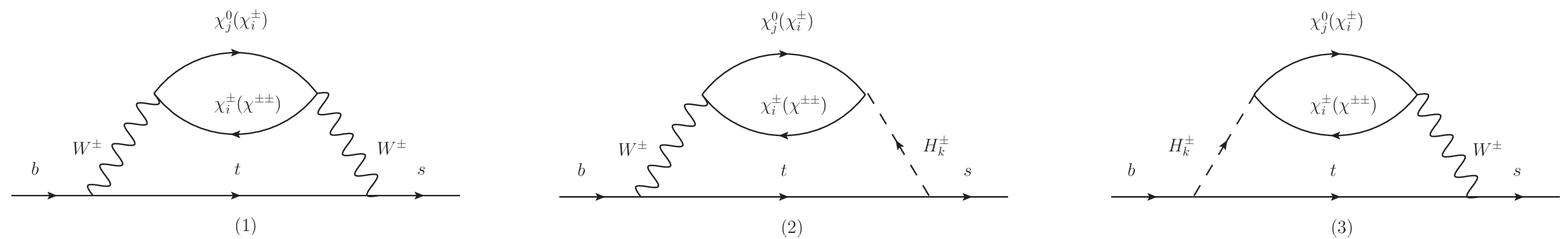}
\vspace{0cm}
\caption[]{The relating two loop diagrams in which a closed heavy fermion loop is attached to virtual $W^\pm$ bosons or $H^\pm$, where a real photon or gluon is attached in all possible ways.}
\label{Bazz diagrams}
\end{figure}

At two loop level, we consider the corrections from closed fermion loop in Fig.~\ref{Bazz diagrams}. In Fig.~\ref{Bazz diagrams}, new defined neutralinos, charginos and doubly-charged chargino in the TNMSSM bring new contributions to $\bar{B}\rightarrow X_s\gamma$ compared to the MSSM. Then, the two loop Wilson coefficients $C_{7,NP}^{WW}(\mu_{EW})$, $C_{8,NP}^{WW}(\mu_{EW})$, $C_{7,NP}^{WH}(\mu_{EW})$ and $C_{8,NP}^{WH}(\mu_{EW})$ from Fig.~\ref{Bazz diagrams} are also collected in Appendix B.

In addition, $C_{8g,NP}(\mu_{EW}),\;C_{8g,NP}^{\prime}(\mu_{EW}),\;C_{8,NP}^{WW}(\mu_{EW}),\;C_{8,NP}^{WH}(\mu_{EW})$ are Wilson coefficients of the process $b\rightarrow sg$ in the TNMSSM, which can make contributions to the process $b\rightarrow s\gamma$ through the QCD running. Similarly, the Wilson coefficients of the process $b\rightarrow sg$ at EW scale can be written as
\begin{eqnarray}
&&C_{8g,NP}(\mu_{EW})=[C_{7,NP}^{(b)}(\mu_{EW})+C_{7,NP}^{(c)}(\mu_{EW})]/Q_u+C_{8,NP}^{WW}(\mu_{EW})+
C_{8,NP}^{WH}(\mu_{EW}), \nonumber\\
&&C_{8g,NP}^{\prime}(\mu_{EW})=C_{8g,NP}(\mu_{EW})(L\leftrightarrow R),
\end{eqnarray}
where $Q_u=2/3$.

Based on Wilson coefficients above, the branching ratio of $\bar{B}\rightarrow X_s\gamma$ in the TNMSSM can be given by
\begin{eqnarray}
&&{\rm Br}(\bar{B}\rightarrow X_s\gamma)=R\Big(|C_{7\gamma}(\mu_b)|^2
+N(E_\gamma)\Big)\;,
\end{eqnarray}
where the overall factor $R=2.47\times10^{-3}$, and the nonperturbative contribution
$N(E_\gamma)=(3.6\pm0.6)\times10^{-3}$\cite{H5}. $C_{7\gamma}(\mu_b)$ is defined as
\begin{eqnarray}
&&C_{7\gamma}(\mu_b)=C_{7\gamma,SM}(\mu_b)+C_{7,NP}(\mu_b),
\end{eqnarray}
where we choose the hadron scale $\mu_b=2.5$ GeV and at NNLO level the SM contribution is $C_{7\gamma,SM}(\mu_b) = -0.3689$~\cite{H5,H6,H7,H8}. The Wilson coefficients for new physics at the bottom quark scale can be written as~\cite{H9,H10}
\begin{eqnarray}
&&C_{7,NP}(\mu_b)\approx0.5696
C_{7,NP}(\mu_{EW})+0.1107 C_{8,NP}(\mu_{EW}),
\end{eqnarray}
where
\begin{eqnarray}
&&C_{7,NP}(\mu_{EW})=C_{7,NP}^{(a)}(\mu_{EW})+C_{7,NP}^{(b)}(\mu_{EW})+
C_{7,NP}^{(c)}(\mu_{EW})+C_{7,NP}^{(d)}(\mu_{EW})+\nonumber\\
&&\qquad\;\qquad\;\qquad\;C_{7,NP}^{\prime(a)}(\mu_{EW})+C_{7,NP}^{\prime(b)}(\mu_{EW})+C_{7,NP}^{\prime
(c)}(\mu_{EW})+C_{7,NP}^{\prime(d)}(\mu_{EW})+\nonumber\\
&&\qquad\;\qquad\;\qquad\;C_{7,NP}^{WW}(\mu_{EW})+C_{7,NP}^{WH}(\mu_{EW}),\nonumber\\
&&C_{8,NP}(\mu_{EW})=C_{8g,NP}(\mu_{EW})+C_{8g,NP}^{\prime}(\mu_{EW})+C_{8,NP}^{WW}(\mu_{EW})+C_{8,NP}^{WH}(\mu_{EW}).
\end{eqnarray}

\section{Rare decay $B_s^0\rightarrow \mu^+\mu^-$ \label{sec4}}
\indent\indent
\begin{figure}
\setlength{\unitlength}{1mm}
\centering
\includegraphics[width=5.5in]{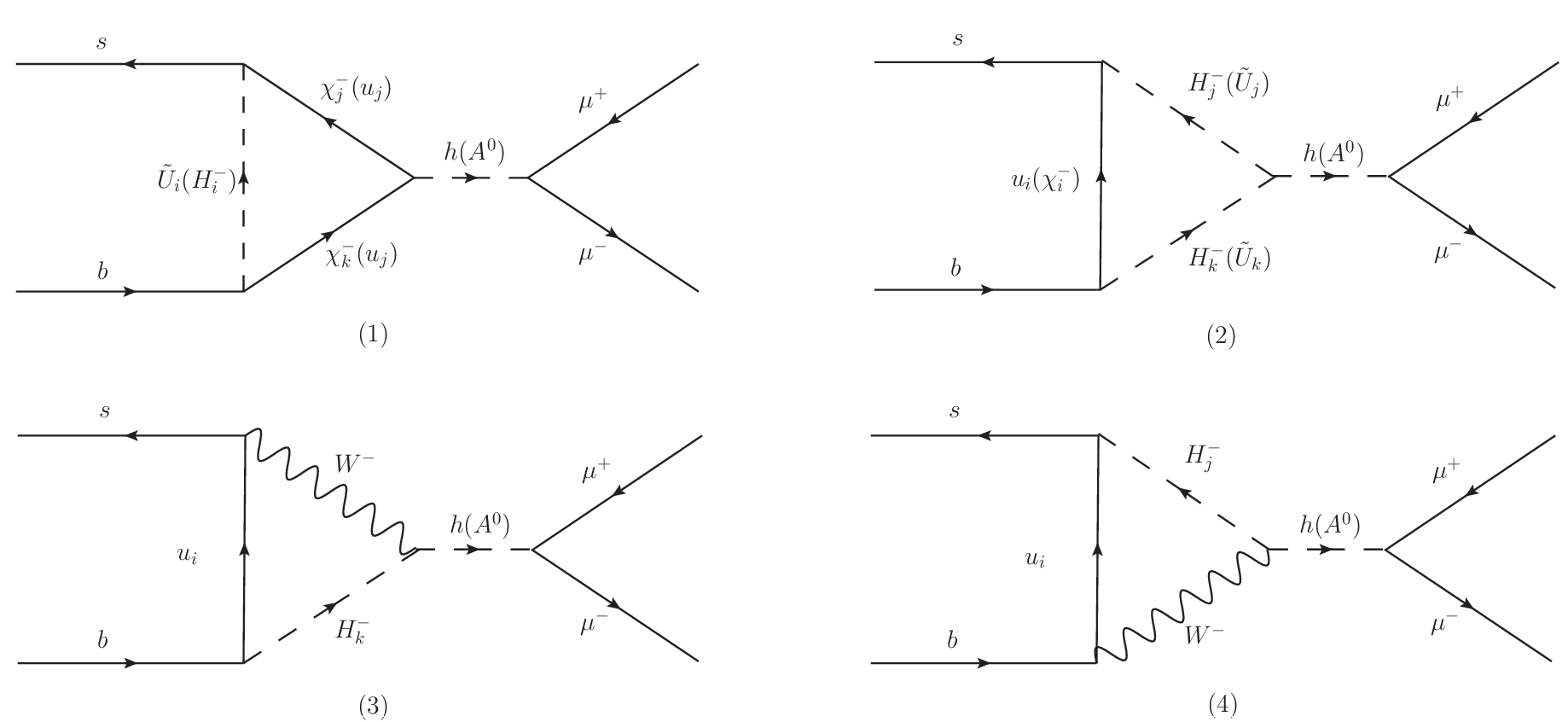}
\vspace{0.5cm}
\includegraphics[width=5.5in]{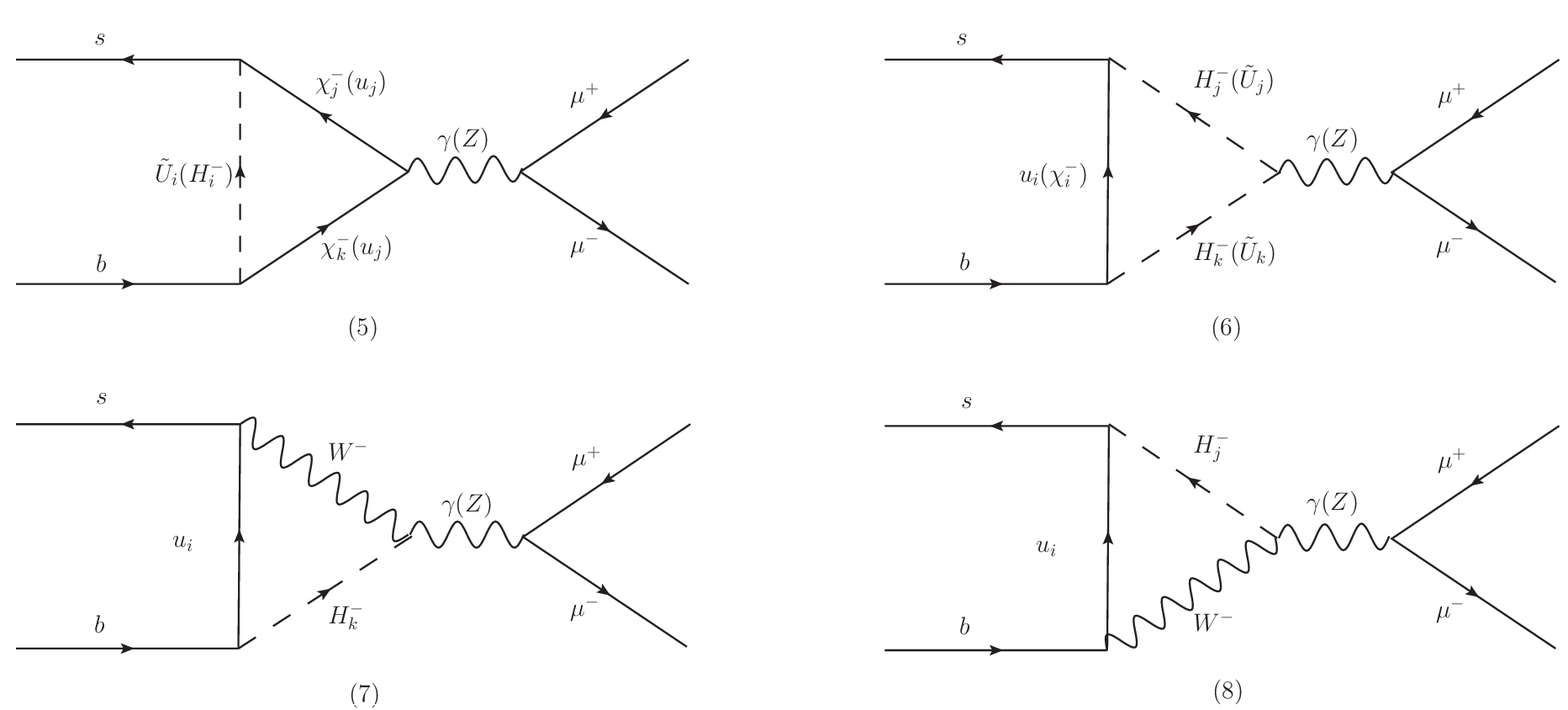}
\vspace{0.5cm}
\includegraphics[width=6in]{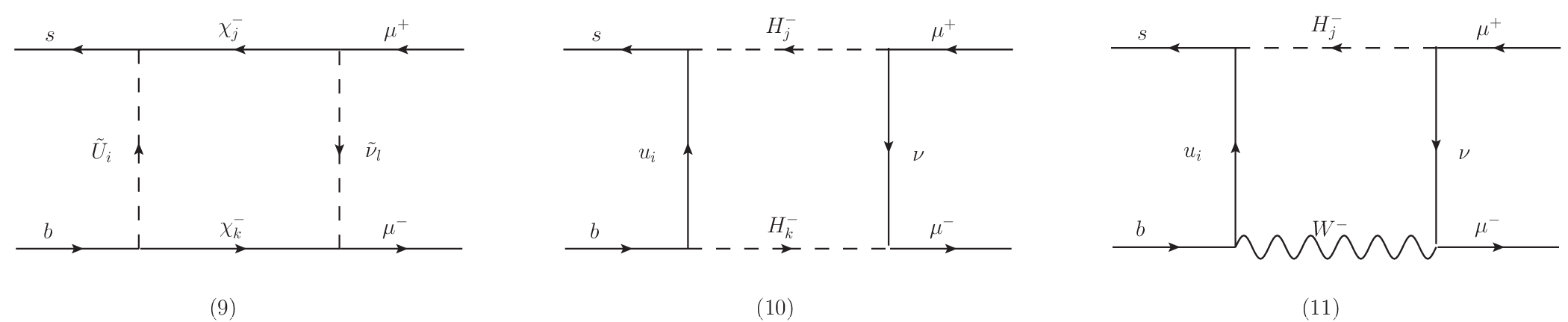}
\vspace{0cm}
\caption[]{The one loop vertex and box diagrams contributing to $B_s^0\rightarrow \mu^+\mu^-$ in the TNMSSM.}
\label{figbmumuOne}
\end{figure}
Fig.~\ref{figbmumuOne} shows the main one loop vertex and box diagrams contributing to the process $B_s^0\rightarrow \mu^+\mu^-$ in the TNMSSM. In Fig.~\ref{figbmumuOne}, newly defined up-squarks and pseudo-scalar Higgs bosons will make new contributions to the branching ratio ${\rm Br}(B_s^0\rightarrow \mu^+\mu^-)$ compared to the MSSM. Considering Fig.~\ref{Bazz diagrams} and Fig.~\ref{figbmumuOne}, the Wilson coefficients corresponding to the process $B_s^0\rightarrow \mu^+\mu^-$ at the EW scale can be written as
\begin{eqnarray}
&&C_{_{S,NP}}(\mu_{_{\rm EW}})=\frac{\sqrt{2}s_{_W}c_{_W}}{4m_be^3V_{ts}^*V_{tb}}\Big[C_{_{S,NP}}^{(1)}(\mu_{_{\rm EW}})+C_{_{S,NP}}^{(2)}(\mu_{_{\rm EW}})+C_{_{S,NP}}^{(3)}(\mu_{_{\rm EW}})+C_{_{S,NP}}^{(4)}(\mu_{_{\rm EW}})\nonumber\\
&&\qquad\;\qquad\;\qquad\;+C_{_{S,NP}}^{(6)}(\mu_{_{\rm EW}})+C_{_{S,NP}}^{(9)}(\mu_{_{\rm EW}})+C_{_{S,NP}}^{(11)}(\mu_{_{\rm EW}})\Big],\nonumber\\
&&C_{_{S,NP}}^\prime(\mu_{_{\rm EW}})=C_{_{S,NP}}(\mu_{_{\rm EW}})(L\leftrightarrow R),\nonumber\\
&&C_{_{P,NP}}(\mu_{_{\rm EW}})=\frac{\sqrt{2}s_{_W}c_{_W}}{4m_be^3V_{ts}^*V_{tb}}\Big[C_{_{P,NP}}^{(1)}(\mu_{_{\rm EW}})+C_{_{P,NP}}^{(2)}(\mu_{_{\rm EW}})+C_{_{P,NP}}^{(3)}(\mu_{_{\rm EW}})+C_{_{P,NP}}^{(4)}(\mu_{_{\rm EW}})\nonumber\\
&&\qquad\;\qquad\;\qquad\;+C_{_{P,NP}}^{(6)}(\mu_{_{\rm EW}})+C_{_{P,NP}}^{(9)}(\mu_{_{\rm EW}})+C_{_{P,NP}}^{(11)}(\mu_{_{\rm EW}})\Big],\nonumber\\
&&C_{_{P,NP}}^\prime(\mu_{_{\rm EW}})=-C_{_{P,NP}}(\mu_{_{\rm EW}})(L\leftrightarrow R),\nonumber\\
&&C_{_{9,NP}}(\mu_{_{\rm EW}})=\frac{\sqrt{2}s_{_W}c_{_W}g_{_s}^2}{64\pi^2e^3V_{ts}^*V_{tb}}\Big[C_{_{9,NP}}^{(5)}(\mu_{_{\rm EW}})+C_{_{9,NP}}^{(6)}(\mu_{_{\rm EW}})+C_{_{9,NP}}^{(7)}(\mu_{_{\rm EW}})+C_{_{9,NP}}^{(8)}(\mu_{_{\rm EW}})\nonumber\\
&&\qquad\;\qquad\;\qquad\;+C_{_{9,NP}}^{(9)}(\mu_{_{\rm EW}})+C_{_{9,NP}}^{(10)}(\mu_{_{\rm EW}})+C_{_{9,NP}}^{WW}(\mu_{_{\rm EW}})\Big]\;,\nonumber\\
&&C_{_{9,NP}}^\prime(\mu_{_{\rm EW}})=C_{_{9,NP}}(\mu_{_{\rm EW}})(L\leftrightarrow R),\nonumber\\
&&C_{_{10,NP}}(\mu_{_{\rm EW}})=\frac{\sqrt{2}s_{_W}c_{_W}g_{_s}^2}{64\pi^2e^3V_{ts}^*V_{tb}}\Big[C_{_{10,NP}}^{(5)}(\mu_{_{\rm EW}})+C_{_{10,NP}}^{(6)}(\mu_{_{\rm EW}})+C_{_{10,NP}}^{(7)}(\mu_{_{\rm EW}})+C_{_{10,NP}}^{(8)}(\mu_{_{\rm EW}})\nonumber\\
&&\qquad\;\qquad\;\qquad\;+C_{_{10,NP}}^{(9)}(\mu_{_{\rm EW}})+C_{_{10,NP}}^{(10)}(\mu_{_{\rm EW}})+C_{_{10,NP}}^{WW}(\mu_{_{\rm EW}})\Big]\;,\nonumber\\
&&C_{_{10,NP}}^\prime(\mu_{_{\rm EW}})=-C_{_{10,NP}}(\mu_{_{\rm EW}})(L\leftrightarrow R).
\label{Wilson-Coefficients1}
\end{eqnarray}
All of the Wilson coefficients calculated above are gauge invariant and the concrete expressions on the right side of Eq.(\ref{Wilson-Coefficients1}) are collected in Appendix B. In addition, the Wilson coefficients can also be evolved from EW scale $\mu_{EW}$ down to hadronic scale $\mu\sim m_b$ by the renormalization group equations. For obtaining hadronic matrix elements conveniently, we define effective coefficients as~\cite{Altmannshofer:2008dz}

\begin{eqnarray}
&&C_7^{eff}=\frac{4\pi}{\alpha_s}C_7-\frac{1}{3}C_3-\frac{4}{9}C_4-\frac{20}{3}C_5-\frac{80}{9}C_6,\nonumber\\
&&C_8^{eff}=\frac{4\pi}{\alpha_s}C_8+C_3-\frac{1}{6}C_4+20C_5-\frac{10}{3}C_6,\nonumber\\
&&C_9^{eff}=\frac{4\pi}{\alpha_s}C_9+Y(q^2),\;\;C_{10}^{eff}=\frac{4\pi}{\alpha_s}C_{10},\nonumber\\
&&{C'}_{7,8,9,10}^{eff}=\frac{4\pi}{\alpha_s}{C'}_{7,8,9,10}.
\end{eqnarray}

\begin{table}
\begin{tabular}{|c|c|c|c|}
\hline
\hline
$C_{_7}^{eff,SM}$    & $C_{_8}^{eff,SM}$    & $C_{_9}^{eff,SM}-Y(q^2)$ & $C_{_{10}}^{eff,SM}$\\
\hline
$-0.304$   & $-0.167$  & $4.211$ & $-4.103$\\
\hline
\hline
\end{tabular}
\caption{At hadronic scale $\mu=m_{_b}\simeq4.65$GeV, Wilson coefficients from the SM to NNLL accuracy. \label{tab1}}
\end{table}
The Wilson coefficients at hadronic energy scale from the SM to Next-to-Next-to-Logarithmic (NNLL) accuracy are shown in Table I. And the renormalization group equations are written as
\begin{eqnarray}
&&\overrightarrow{C}_{_{NP}}(\mu)=\widehat{U}(\mu,\mu_0)\overrightarrow{C}_{_{NP}}(\mu_0)
\;,\nonumber\\
&&\overrightarrow{C^\prime}_{_{NP}}(\mu)=\widehat{U^\prime}(\mu,\mu_0)
\overrightarrow{C^\prime}_{_{NP}}(\mu_0)
\label{evaluation1}
\end{eqnarray}
with
\begin{eqnarray}
&&\overrightarrow{C}_{_{NP}}^{T}=\Big(C_{_{1,NP}},\;\cdots,\;C_{_{6,NP}},
C_{_{7,NP}}^{eff},\;C_{_{8,NP}}^{eff},\;C_{_{9,NP}}^{eff}-Y(q^2),\;
C_{_{10,NP}}^{eff}\Big)
\;,\nonumber\\
&&\overrightarrow{C}_{_{NP}}^{\prime,\;T}=\Big(C_{_{7,NP}}^{\prime,\;eff},\;
C_{_{8,NP}}^{\prime,\;eff},\;C_{_{9,NP}}^{\prime,\;eff},\;
C_{_{10,NP}}^{\prime,\;eff}\Big)\;,
\label{evaluation2}
\end{eqnarray}
and
\begin{eqnarray}
&&\widehat{U}(\mu,\mu_0)\simeq1-\Big[{1\over2\beta_0}\ln{\alpha_{_s}(\mu)\over
\alpha_{_s}(\mu_0)}\Big]\widehat{\gamma}^{(0)T}
\;,\nonumber\\
&&\widehat{U^\prime}(\mu,\mu_0)\simeq1-\Big[{1\over2\beta_0}\ln{\alpha_{_s}(\mu)\over
\alpha_{_s}(\mu_0)}\Big]\widehat{\gamma^\prime}^{(0)T}\;,
\label{evaluation3}
\end{eqnarray}
where Ref.~\cite{Gambino1} provided the anomalous dimension matrices as
\begin{eqnarray}
&&\widehat{\gamma}^{(0)}=\left(\begin{array}{cccccccccc}
-4&{8\over3}&0&-{2\over9}&0&0&-{208\over243}&{173\over162}&-{2272\over729}&0\\
12&0&0&{4\over3}&0&0&{416\over81}&{70\over27}&{1952\over243}&0\\
0&0&0&-{52\over3}&0&2&-{176\over81}&{14\over27}&-{6752\over243}&0\\
0&0&-{40\over9}&-{100\over9}&{4\over9}&{5\over6}&-{152\over243}&-{587\over162}&-{2192\over729}&0\\
0&0&0&-{256\over3}&0&20&-{6272\over81}&{6596\over27}&-{84032\over243}&0\\
0&0&-{256\over9}&{56\over9}&{40\over9}&-{2\over3}&{4624\over243}&{4772\over81}&-{37856\over729}&0\\
0&0&0&0&0&0&{32\over3}&0&0&0\\
0&0&0&0&0&0&-{32\over9}&{28\over3}&0&0\\
0&0&0&0&0&0&0&0&0&0\\
0&0&0&0&0&0&0&0&0&0\\
\end{array}\right)
\;,\nonumber\\
&&\widehat{\gamma^\prime}^{(0)}=\left(\begin{array}{cccc}
{32\over3}&0&0&0\\
-{32\over9}&{28\over3}&0&0\\
0&0&0&0\\0&0&0&0\\
\end{array}\right)\;.
\label{ADM1}
\end{eqnarray}

Based on the Wilson coefficients above, the branching ratio of $B_s^0\rightarrow \mu^+\mu^-$ can be given by
\begin{eqnarray}
&&{\rm Br}(B_s^0\rightarrow\mu^+\mu^-)=\frac{\tau_{B_s^0}}{16\pi}\frac{|\mathcal{M}_s|^2}{M_{B_s^0}}\sqrt{1-\frac{4m_{\mu}^2}{M_{B_s^0}^2}},
\end{eqnarray}
where $M_{B_s^0}=5.367 \rm GeV$ denotes the mass of neutral meson $B_s^0$ and $\tau_{B_s^0}=1.466(31){\rm ps}$ denotes its life time. Moreover, the squared amplitude can be written as
\begin{eqnarray}
&&|\mathcal{M}_s|^2=16G_F^2|V_{tb}V_{ts}^*|^2M_{B_s^0}^2\Big[|F_S^s|^2+|F_P^s+2m_{\mu}F_A^s|^2\Big],
\end{eqnarray}
with
\begin{eqnarray}
&&F_S^s=\frac{\alpha_{EW}(\mu_b)}{8\pi}\frac{m_b M_{B_s^0}^2}{m_b+m_s}f_{B_s^0}(C_S-C_S'),\\
&&F_P^s=\frac{\alpha_{EW}(\mu_b)}{8\pi}\frac{m_b M_{B_s^0}^2}{m_b+m_s}f_{B_s^0}(C_P-C_P'),\\
&&F_A^s=\frac{\alpha_{EW}(\mu_b)}{8\pi}f_{B_s^0}\Big[C_{10}^{eff}(\mu_b)-C_{10}^{\prime eff}(\mu_b)\Big],
\end{eqnarray}
where $f_{B_s^0}=(227\pm8){\rm MeV}$ denote the decay constants.

\section{Numerical analyses\label{sec5}}
\indent\indent
This section provides the numerical discussion of the branching ratios of B meson rare decays $\bar B\rightarrow X_s\gamma$ and $B_s^0\rightarrow \mu^+\mu^-$ by considering the latest multiple experimental constraints of particles. It includes that the SM-like Higgs mass $m_h$ is keeped around $125.25\;{\rm GeV}$, the neutralino mass is limited to more than $116\;{\rm GeV}$, the chargino mass is limited to more than $1100\;{\rm GeV}$, the slepton mass is limited to more than $700\;{\rm GeV}$ and the squark mass is maintained at the TeV order of magnitude \cite{17,CMS.PLB,ATLAS.PLB,PC,MV,MC,FW,MC1,ME}. Additionally, the first two generations of squarks are strongly constrained by direct searches at the LHC \cite{ATLAS.PRD,CMS.JHEP}. However, compared to the previous two generations, the mass of the third generation squark which can affect the mass of SM-like Higgs do not suffer strong constraints. Therefore we take $m_{\tilde{q}}=m_{\tilde{u}}=diag(2, 2, m_{\tilde t})\;{\rm TeV}$, and the discussion about the observed Higgs signal in Ref.\cite{add1} limits $m_{\tilde t}\gtrsim1.5\;{\rm TeV}$. For simplicity, we also choose $T_{u_{1,2}}=Y_{u_{1,2}} A_{u_{1,2}}$, $A_{u_{1,2}}=1\;{\rm TeV}$. As a key parameter, $T_{u_3}=A_t$ affects the SM-like Higgs mass and the numerical calculation obviously. In order to obtain reasonable numerical results, we need to find some sensitive parameters for discussion. Similarly to the MSSM \cite{MH}, the new physics contributions to the branching ratios ${\rm Br}(\bar B\rightarrow X_s\gamma)$ and ${\rm Br}(B_s^0\rightarrow \mu^+\mu^-)$ are also depended on $\tan\beta$, $A_t$ and singly-charged Higgs mass. Moreover, according to their great impacts on singly-charged Higgs mass, neutralino mass and chargino mass, we found three other new parameters $\kappa$, $\lambda$ and $T_{\lambda}$ that can affect ${\rm Br}(\bar B\rightarrow X_s\gamma)$ and ${\rm Br}(B_s^0\rightarrow \mu^+\mu^-)$ in the TNMSSM. We will plot the relational and scatter diagrams and explore the effects of these parameters on the branching ratios and the allowed ranges between $\lambda$, $\kappa$ and $T_{\lambda}$.

By considering the experimental constraints described above, we adopt the parameters in the following numerical calculation as
\begin{eqnarray}
&&\chi_d=\chi_u=0.1,\;\Lambda_T=0.6,\;\tan\beta'=5,\;M_1=1\;{\rm TeV},\;m_{\tilde t}=1.6\;{\rm TeV},\nonumber\\
&&M_2=\mu=1.2\;{\rm TeV},\;T_{\chi_d}=T_{\chi_u}=-500\;{\rm GeV},\;T_{\kappa}=-300\;{\rm GeV},\;T_{\Lambda_T}=100\;{\rm GeV}.
\label{parameter space}
\end{eqnarray}

\begin{figure}
\setlength{\unitlength}{1mm}
\centering
\includegraphics[width=3.1in]{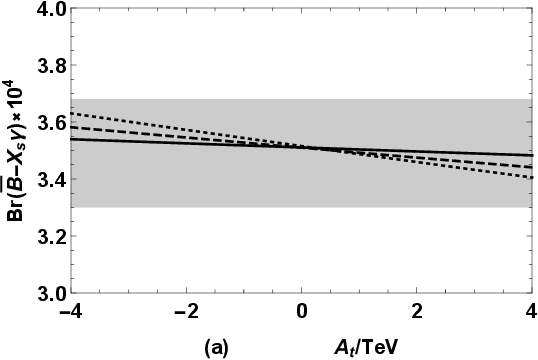}%
\vspace{0.5cm}
\includegraphics[width=3.1in]{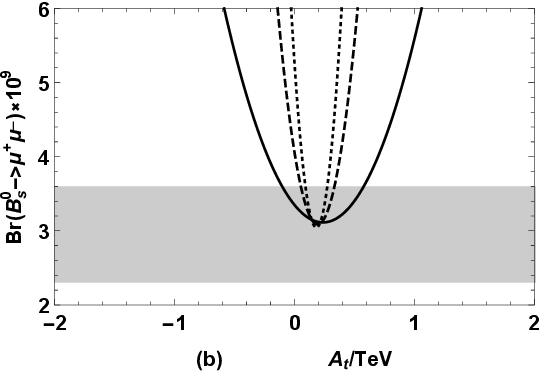}%
\vspace{0cm}
\caption[]{(a) ${\rm Br}(\bar B\rightarrow X_s\gamma)$ and (b) ${\rm Br}(B_s^0\rightarrow \mu^+\mu^-)$ versus $A_t$ for $\tan\beta=10\;({\rm solid \;line}),\;\tan\beta=25\;({\rm dashed \;line}),\;\tan\beta=40\;({\rm dotted \;line})$, where the gray area denotes the experimental 1$\sigma$ interval.}
\label{figAt}
\end{figure}

To illustrate the effects of $A_t$ and $\tan\beta$ on the branching ratios, we take $\lambda=0.4,\;\kappa=0.6,\;T_{\lambda}=100\;{\rm GeV}$ and plot the graph of ${\rm Br}(\bar B\rightarrow X_s\gamma)$ and ${\rm Br}(B_s^0\rightarrow \mu^+\mu^-)$ varying with $A_t$ in Fig.~\ref{figAt} for $\tan\beta=10\;({\rm solid \;line}),\;\tan\beta=25\;({\rm dashed \;line}),\;\tan\beta=40\;({\rm dotted \;line})$. The gray area denotes the experimental 1$\sigma$ bounds in Eq.(\ref{experimental data}). Fig.~\ref{figAt}(a) shows that ${\rm Br}(\bar B\rightarrow X_s\gamma)$ decreases with the increasing of $A_t$ and the slope of evolution is steeper as $\tan\beta$ is bigger. In Fig.~\ref{figAt}(b), we can see that the variation relationship between $A_t$ and ${\rm Br}(B_s^0\rightarrow \mu^+\mu^-)$ is almost the shape of parabolic. Moreover, when $\tan\beta$ is bigger, $A_t$ is limited strongly in the smaller range by the experimental data on ${\rm Br}(B_s^0\rightarrow \mu^+\mu^-)$. It will be seen from this that the new physics can provide the considerable contributions to ${\rm Br}(B_s^0\rightarrow \mu^+\mu^-)$ for large $\tan\beta$ and $A_t$.

\begin{figure}
\setlength{\unitlength}{1mm}
\centering
\includegraphics[width=3.1in]{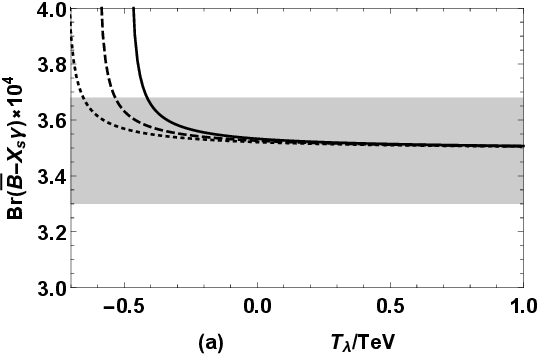}%
\vspace{0.5cm}
\includegraphics[width=3.1in]{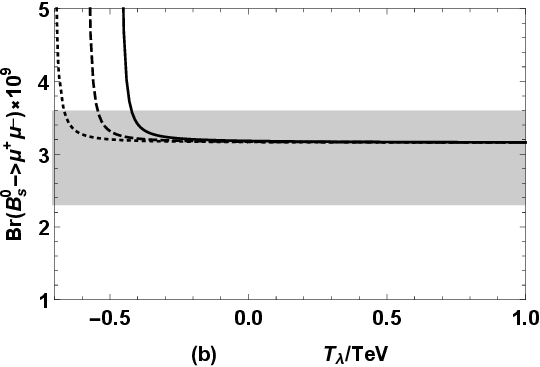}%
\vspace{0cm}
\caption[]{(a) ${\rm Br}(\bar B\rightarrow X_s\gamma)$ and (b) ${\rm Br}(B_s^0\rightarrow \mu^+\mu^-)$ versus $T_{\lambda}$ for $\kappa=0.4\;({\rm solid \;line}),\;\kappa=0.5\;({\rm dashed \;line}),\;\kappa=0.6\;({\rm dotted \;line})$ when $\lambda=0.4$. The gray area denotes the experimental 1$\sigma$ interval.}
\label{figTlambda}
\end{figure}

\begin{figure}
\setlength{\unitlength}{1mm}
\centering
\includegraphics[width=3.1in]{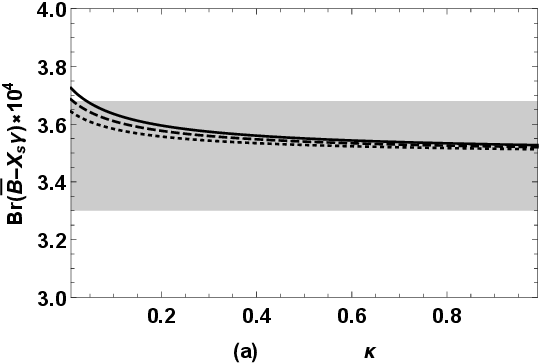}%
\vspace{0.5cm}
\includegraphics[width=3.1in]{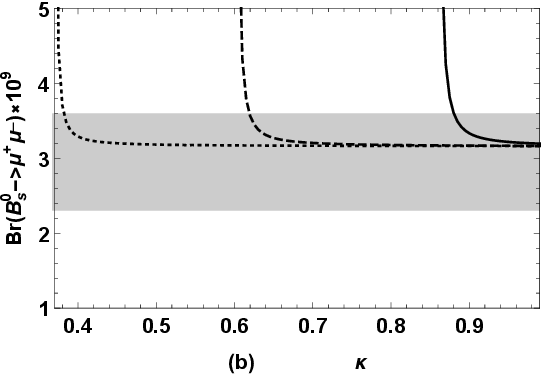}%
\vspace{0cm}
\caption[]{(a) ${\rm Br}(\bar B\rightarrow X_s\gamma)$ and (b) ${\rm Br}(B_s^0\rightarrow \mu^+\mu^-)$ versus $\kappa$ for $\lambda=0.9\;({\rm solid \;line}),\;\lambda=0.7\;({\rm dashed \;line}),\;\lambda=0.5\;({\rm dotted \;line})$ when $T_{\lambda}=100\;{\rm GeV}$. The gray area denotes the experimental 1$\sigma$ interval.}
\label{figkappa}
\end{figure}

To see the effects of $\lambda,\;\kappa,$ and $T_{\lambda}$ on the branching ratios, we first take $\lambda=0.4,\;\tan\beta=5,\;A_t=0.5\;{\rm TeV}$ and plot the graph of ${\rm Br}(\bar B\rightarrow X_s\gamma)$ and ${\rm Br}(B_s^0\rightarrow \mu^+\mu^-)$ varying with $T_{\lambda}$ in Fig.~\ref{figTlambda}, for $\kappa=0.4\;({\rm solid \;line}),\;\kappa=0.5\;({\rm dashed \;line}),\;\kappa=0.6\;({\rm dotted \;line})$, respectively. Then, we take $T_{\lambda}=100\;{\rm GeV},\;\tan\beta=5,\;A_t=0.5\;{\rm TeV}$ and plot the graph of ${\rm Br}(\bar B\rightarrow X_s\gamma)$ and ${\rm Br}(B_s^0\rightarrow \mu^+\mu^-)$ varying with $\kappa$ in Fig.~\ref{figkappa}, for $\lambda=0.9\;({\rm solid \;line}),\;\lambda=0.7\;({\rm dashed \;line}),\;\lambda=0.5\;({\rm dotted \;line})$, respectively. The gray area denotes the experimental 1$\sigma$ bounds. Fig.~\ref{figTlambda} shows that ${\rm Br}(\bar B\rightarrow X_s\gamma)$ and ${\rm Br}(B_s^0\rightarrow \mu^+\mu^-)$ increase with the decreasing of $T_{\lambda}$ when $T_{\lambda}$ is negative and the three lines mix together for large $T_{\lambda}$. Thus it can be seen that the smaller $\kappa$ is taken, the smaller range of $T_{\lambda}$ is limited and large $T_{\lambda}$ can get rid of the influence of $\kappa$. In Fig.~\ref{figkappa}(a), ${\rm Br}(\bar B\rightarrow X_s\gamma)$ decreases gradually with the increasing of $\kappa$ and $\kappa$ is less affected by $\lambda$. However, from Fig.~\ref{figkappa}(b), ${\rm Br}(B_s^0\rightarrow \mu^+\mu^-)$ decreases dramatically near $\kappa=0.4$ for $\lambda=0.5$, $\kappa=0.6$ for $\lambda=0.7$ and $\kappa=0.9$ for $\lambda=0.9$ respectively. It indicates that $\lambda$ has strong limitations on $\kappa$ and the limitations will be discussed in the following by drawing scatter plots.

\begin{figure}
\setlength{\unitlength}{1mm}
\centering
\includegraphics[width=3.1in]{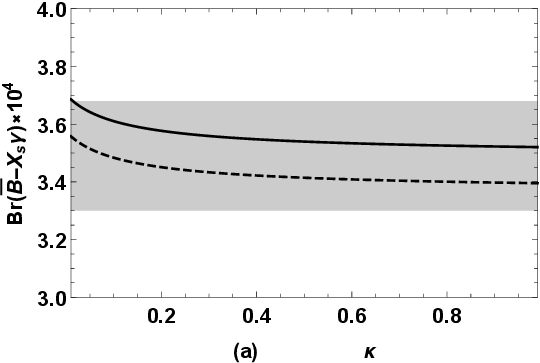}%
\vspace{0.5cm}
\includegraphics[width=3.1in]{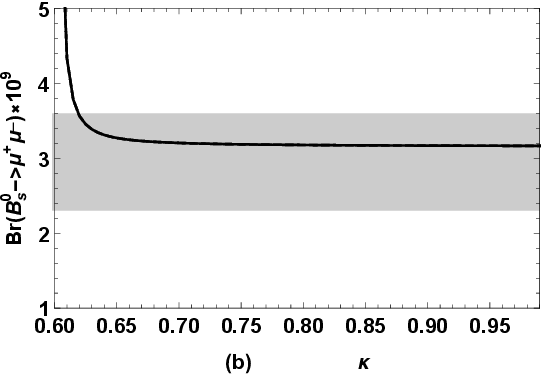}%
\vspace{0cm}
\caption[]{The comparison graph of two loop (solid lines) result and one loop (dashed lines) result when $\lambda=0.7,\;T_{\lambda}=100\;{\rm GeV}$, where the gray area denotes the experimental 1$\sigma$ interval.}
\label{figcompared}
\end{figure}

Meanwhile, for comparing and reflecting the specific differences between one loop and two loop corrections to $\bar B\rightarrow X_s\gamma$ and $B_s^0\rightarrow \mu^+\mu^-$, we take $T_{\lambda}=100\;{\rm GeV}$, $\lambda=0.7$ and plot the graph of ${\rm Br}(\bar B\rightarrow X_s\gamma)$ and ${\rm Br}(B_s^0\rightarrow \mu^+\mu^-)$ varying with $\kappa$ in Fig.~\ref{figcompared}. The solid and dashed line denote two loop and one loop predictions respectively. Fig.~\ref{figcompared}(a) shows that, the relative corrections from two loop diagrams to one loop corrections of ${\rm Br}(\bar B\rightarrow X_s\gamma)$ can reach around $4\%$, which can produce a more precise prediction on the process $\bar B\rightarrow X_s\gamma$. In Fig.~\ref{figcompared}(b), we can see that the two lines almost overlap which shows that the two loop corrections are negligible compared with one loop corrections. Therefore, in the analysis of the numerical calculations above, we always use the more precise two loop predictions.

\begin{figure}
\setlength{\unitlength}{1mm}
\centering
\includegraphics[width=3.1in]{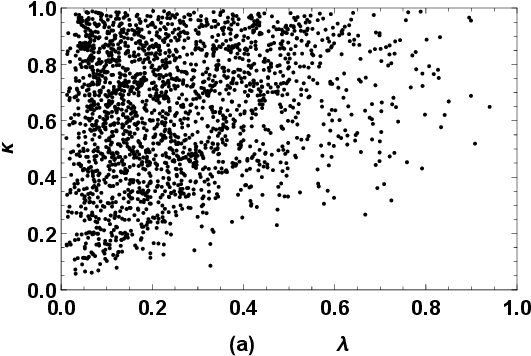}%
\vspace{0.5cm}
\includegraphics[width=3.1in]{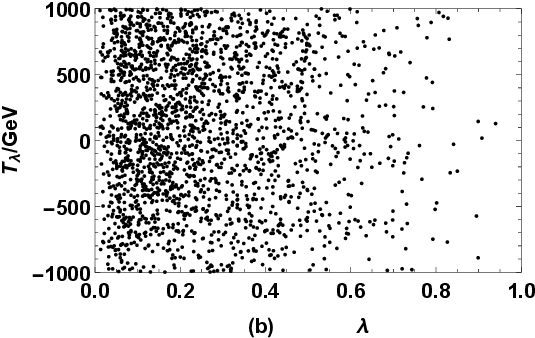}%
\vspace{0.5cm}
\includegraphics[width=3.5in]{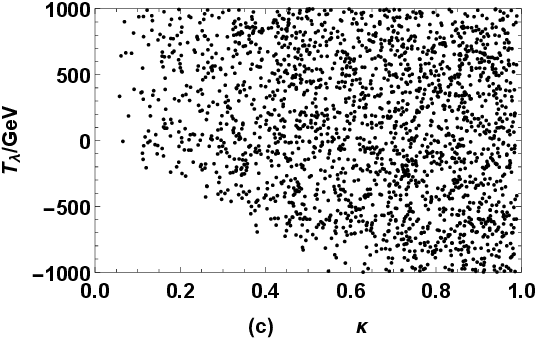}%
\vspace{0cm}
\caption[]{Scanning the parameter space in Eq.(\ref{scan}) and keeping ${\rm Br}(\bar B\rightarrow X_s\gamma)$, ${\rm Br}(B_s^0\rightarrow \mu^+\mu^-)$ in the experimental 1$\sigma$ interval, the allowed ranges of $\kappa-\lambda$ (a), $T_{\lambda}-\lambda$ (b), $T_{\lambda}-\kappa$ (c) are plotted.}
\label{figscatter}
\end{figure}

Now, for revealing how $\lambda,\;\kappa,$ and $T_{\lambda}$ be constrained by the experimental measurements of B meson rare decays, we scan the sensitive parameters under the consideration of the experimental constraints above and ${\rm Br}(\bar B\rightarrow X_s\gamma)$, ${\rm Br}(B_s^0\rightarrow \mu^+\mu^-)$ within one standard deviation. The random ranges of input parameters are as follows:
\begin{eqnarray}
&&\tan\beta=(1,40),\;\lambda=(0.01,0.99),\;\kappa=(0.01,0.99),\nonumber\\
&&\mu=(1.1,1.5)\;{\rm TeV},\;T_{\lambda}=(-1,1)\;{\rm TeV},\;A_t=(-4,4)\;{\rm TeV}.
\label{scan}
\end{eqnarray}
Then, we plot the allowed ranges of $\kappa$ versus $\lambda$, $T_{\lambda}$ versus $\lambda$ and $\kappa$ versus $T_{\lambda}$ in Fig.~\ref{figscatter}. Fig.~\ref{figscatter}(a) shows that the vast majority of points are concentrated in the areas $\kappa>\lambda$ and the number of points gradually decreases as $\kappa<\lambda$. In Fig.~\ref{figscatter}(b), we can find that the density of points decreases with the increasing of $\lambda$ and this phenomenon is more obvious when $\lambda>0.6$. As shown in Fig.~\ref{figscatter}(c), the negative range of $T_{\lambda}$ is gradually limited when $\kappa<0.6$. In conclusion, maintaining ${\rm Br}(\bar B\rightarrow X_s\gamma)$ and ${\rm Br}(B_s^0\rightarrow \mu^+\mu^-)$ within the experimental 1$\sigma$ interval prefers $\lambda,\;\kappa$ in the range $\lambda<0.6$, $\kappa>0.6$ and positive $T_{\lambda}$.

\section{Summary\label{sec6}}
\indent\indent
B meson rare decays are sensitive to the searching on new physics beyond the SM. In this paper, we investigate the two loop electroweak corrections to the processes $\bar B\rightarrow X_s\gamma$ and $B_s^0\rightarrow \mu^+\mu^-$ in the TNMSSM which extends the MSSM with two triplets and one singlet. Under the consideration of the constraints from the observed Higgs signal and the updated experimental data of the branching ratios, the numerical results indicate that the corrections from two loop diagrams to the process $\bar B\rightarrow X_s\gamma$ in the TNMSSM can reach around $4\%$, which can produce a more precise theoretical prediction. Moreover, the new physics effects in the TNMSSM can fit the experimental data for the rare decays $\bar B\rightarrow X_s\gamma$ and $B_s^0\rightarrow \mu^+\mu^-$, and the corresponding parameter space is limited strictly by considering ${\rm Br}(\bar B\rightarrow X_s\gamma)$ and ${\rm Br}(B_s^0\rightarrow \mu^+\mu^-)$ in the experimental 1$\sigma$ intervals. The new parameters $\lambda,\;\kappa,\;T_{\lambda}$ in the TNMSSM have great impacts on the theoretical predictions of ${\rm Br}(\bar B\rightarrow X_s\gamma)$, ${\rm Br}(B_s^0\rightarrow \mu^+\mu^-)$ and maintaining ${\rm Br}(\bar B\rightarrow X_s\gamma)$ and ${\rm Br}(B_s^0\rightarrow \mu^+\mu^-)$ within the experimental 1$\sigma$ interval prefers $\lambda,\;\kappa$ in the range $\lambda<0.6$, $\kappa>0.6$ and positive $T_{\lambda}$.

\begin{acknowledgments}
\indent\indent
Our deepest gratitude goes to professor Tai-Fu Feng for his sincere guidance on this research. The work has been supported by the National Natural Science Foundation of China (NNSFC) with Grants No. 12075074, No. 12235008, No. 11535002, No. 11705045, Hebei Natural Science Foundation for Distinguished Young Scholars with Grant No. A2022201017, Natural Science Foundation of Guangxi Autonomous Region with Grant No. 2022GXNSFDA035068, and the youth top-notch talent support program of the Hebei Province.
\end{acknowledgments}

\appendix

\section{The corresponding matrix elements}
The matrix elements of the squared mass matrix for singly-charged Higgs can are given by
\begin{eqnarray}
&&m_{H_d^- H_d^{-,\ast}}=\frac{\sqrt2}{2} T_{\lambda} \tan\beta v_S -\sqrt2 T_{\chi_d} v_T +\frac{1}{4} g_2^2 (\tan^2\beta v_d^2 +2 v_T^2 -2 v_{\bar T}^2) -\frac{1}{2} \lambda^2 \tan^2\beta v_d^2 \nonumber\\
&&\hspace{1.4cm}
+\Lambda_T \chi_d v_S v_{\bar T} -2 {\chi_d}^2 v_T^2 +\frac{1}{2} \lambda \tan\beta (\kappa v_S^2 -\Lambda_T v_T v_{\bar T} +2 v_S (\chi_d v_T +\chi_u v_{\bar T})), \nonumber\\
&&m_{H_u^{+,\ast} H_u^+}=\frac{1}{4 \tan\beta} (2 \sqrt2 T_{\lambda} v_S+g_2^2 \tan\beta (v_d^2 -2 v_T^2 +2 v_{\bar T}^2) -2 \tan\beta (2 \sqrt2 T_{\chi_u} v_{\bar T} +\lambda^2 v_d^2 \nonumber\\
&&\hspace{1.4cm}
-2 \Lambda_T \chi_u v_S v_T +4 \chi_u^2 v_{\bar T}^2) +2 \lambda (\kappa v_S^2 -\Lambda_T v_T v_{\bar T} +2 (\chi_d v_T +\chi_u v_{\bar T}) v_S)), \nonumber\\
&&m_{{\bar T}^- {\bar T}^{-,\ast}}=\frac{1}{4} (2 \sqrt2 T_{\Lambda_T} \tan\beta' v_S -g_2^2 v_d^2 +2 g_2^2 v_T^2 +2 \kappa \Lambda_T \tan\beta' v_S^2 -2 \Lambda_T^2 v_T^2 +\frac{2 \Lambda_T \chi_d v_d^2 v_S}{v_{\bar T}} \nonumber\\
&&\hspace{1.4cm}
-2 \tan\beta \lambda v_d^2 (\Lambda_T \tan\beta' -\frac{2 \chi_u v_S}{v_{\bar T}}) +\tan^2\beta v_d^2 (-\frac{2 \sqrt2 T_{\chi_u}}{v_{\bar T}} +g_2^2 -4 \chi_u^2)), \nonumber\\
&&m_{T^{+,\ast} T^+}=\frac{1}{4} (-\frac{2 \sqrt2 T_{\chi_d} v_d^2}{v_T} +g_2^2 (v_d^2 -\tan^2\beta v_d^2 +2 v_{\bar T}^2) +2 (\frac{\sqrt2 T_{\Lambda_T} v_S}{\tan\beta'} \nonumber\\
&&\hspace{1.4cm}
+\frac{\kappa \Lambda_T v_S^2}{\tan\beta'} -\frac{\lambda \Lambda_T \tan\beta v_d^2}{\tan\beta'} -\Lambda_T^2 v_{\bar T}^2 -2 \chi_d^2 v_d^2 +\frac{\tan\beta v_d^2 v_S}{v_T} (2 \lambda \chi_d +\tan\beta \Lambda_T \chi_u))), \nonumber\\
&&m^{\ast}_{H_u^{+,\ast} H_d^{-,\ast}}=m_{H_d^- H_u^+}=\frac{1}{4} (g_2^2 \tan\beta v_d^2 +2 \sqrt2 T_{\lambda} v_S -2 \lambda (-\kappa v_S^2 +\lambda \tan\beta v_d^2 +\Lambda_T v_T v_{\bar T})), \nonumber\\
&&m^{\ast}_{{\bar T}^- H_d^{-,\ast}}=m_{H_d^- {\bar T}^{-,\ast}}=\frac{1}{2 \sqrt2} v_d (g_2^2 v_{\bar T}-2 (\Lambda_T \chi_d+\lambda \chi_u \tan\beta) v_S), \nonumber\\
&&m^{\ast}_{T^{+,\ast} H_d^{-,\ast}}=m_{H_d^- T^+}=\frac{1}{4} v_d (-4 T_{\chi_d}+\sqrt2 (g_2^2 v_T+2 \chi_d (\lambda \tan\beta v_S -2 \chi_d v_T))), \nonumber\\
&&m^{\ast}_{{\bar T}^- H_u^+}=m_{H_u^{+,\ast} {\bar T}^{-,\ast}}=\frac{1}{4} v_d (-4 \tan\beta T_{\chi_u} +2 \sqrt2 \lambda \chi_u v_S +\sqrt2 \tan\beta v_{\bar T} (g_2^2 -4 \chi_u^2)), \nonumber\\
&&m^{\ast}_{T^{+,\ast} H_u^+}=m_{H_u^{+,\ast} T^+}=\frac{v_d}{2 \sqrt2} (g_2^2 \tan\beta v_T -2 v_S (\lambda \chi_d +\tan\beta \Lambda_T \chi_u)), \nonumber\\
&&m^{\ast}_{T^{+,\ast} {\bar T}^{-,\ast}}=m_{{\bar T}^- T^+}=\frac{1}{2} (\sqrt2 T_{\Lambda_T} v_S +g_2^2 v_T v_{\bar T} -\Lambda_T (-\kappa v_S^2 +\lambda \tan\beta v_d^2 +\Lambda_T v_T v_{\bar T})),
\end{eqnarray}
where $\tan\beta=\frac{v_u}{v_d}$ and $\tan\beta'=\frac{v_T}{v_{\bar T}}$.

The matrix elements of the squared mass matrix for CP-even Higgs are written as
\begin{eqnarray}
&&m_{H_d^0 H_d^0}=\frac{1}{4} ((g^2 +8 \chi_d^2) v_d^2+2 \tan\beta v_S N -2 \lambda \Lambda_T \tan\beta \tan\beta' v_{\bar T}^2), \nonumber\\
&&m_{H_u^0 H_u^0}=\frac{1}{4} ((g^2 +8 \chi_u^2) v_d^2+\frac{2 v_S}{\tan\beta} N -2 \lambda \Lambda_T \frac{\tan\beta'}{\tan\beta} v_{\bar T}^2), \nonumber\\
&&m_{H_u^0 H_d^0}=\frac{1}{4} ((-g^2 +4 \lambda^2) \tan\beta v_d^2-2 v_S N +2 \lambda \Lambda_T \tan\beta' v_{\bar T}^2), \nonumber\\
&&m_{S S}=\frac{1}{2} (\frac{\tan\beta v_d^2}{v_S} (x+\sqrt2 T_{\lambda}+2 n)+\frac{1}{v_S} (\Lambda_T \chi_d v_d^2 v_{\bar T}+\sqrt2 T_{\Lambda_T} \tan\beta' v_{\bar T}^2) \nonumber\\
&&\hspace{1.4cm}
+\sqrt2 T_{\kappa} v_S+4 \kappa^2 v_S^2), \nonumber\\
&&m_{S H_d^0}=\frac{-1}{2} v_d (\sqrt2 \tan\beta T_{\lambda}-2 \lambda^2 v_S+2 \tan\beta n+2 \tan\beta \lambda \kappa v_S+2 \Lambda_T \chi_d v_{\bar T}), \nonumber\\
&&m_{S H_u^0}=-\frac{T_{\lambda} v_d}{\sqrt2}-v_d (\lambda v_S (\kappa-\lambda \tan\beta)+n+x), \nonumber\\
&&m_{T^0 T^0}=g^2 \tan^2\beta' v_{\bar T}^2+\frac{1}{2 \tan\beta' v_{\bar T}} (\sqrt2 T_{\Lambda_T} v_S v_{\bar T}-\sqrt2 T_{\chi_d} v_d^2+\kappa \Lambda_T v_S^2 v_{\bar T} \nonumber\\
&&\hspace{1.4cm}
+\tan\beta v_d^2 (2 \lambda \chi_d v_S+\tan\beta \Lambda_T \chi_u v_S-\lambda \Lambda_T v_{\bar T})), \nonumber\\
&&m_{H_d^0 T^0}=\frac{v_d}{2} (2 \sqrt2 T_{\chi_d}-g^2 \tan\beta' v_{\bar T}+\lambda \tan\beta (\Lambda_T v_{\bar T}-2 \chi_d v_S)+8 \tan\beta' \chi_d^2 v_{\bar T}), \nonumber\\
&&m_{H_u^0 T^0}=\frac{v_d}{2} (g^2 \tan\beta \tan\beta' v_{\bar T}+\lambda \Lambda_T v_{\bar T}-2 \lambda \chi_d v_S-2 \tan\beta \Lambda_T \chi_u v_S), \nonumber\\
&&m_{S T^0}=-\frac{T_{\Lambda_T} v_{\bar T}}{\sqrt2}+ \Lambda_T v_S v_{\bar T} (-\kappa+\tan\beta' \Lambda_T) -\frac{1}{2} \tan\beta v_d^2 (2 \lambda \chi_d +\tan\beta \Lambda_T \chi_u), \nonumber\\
&&m_{{\bar T}^0 {\bar T}^0}=g^2 v_{\bar T}^2+\frac{1}{2} \tan\beta' v_S (\sqrt2 T_{\Lambda_T}+\kappa \Lambda_T v_S)+\frac{1}{2 v_{\bar T}} (-\sqrt2 \tan^2\beta T_{\chi_u} v_d^2 \nonumber\\
&&\hspace{1.4cm}
+\Lambda_T \chi_d v_d^2 v_S+\lambda \tan\beta v_d^2 (-\tan\beta' \Lambda_T v_{\bar T}+2 \chi_u v_S)), \nonumber\\
&&m_{H_d^0 {\bar T}^0}=\frac{1}{2} v_d (g^2 v_{\bar T}+\lambda \Lambda_T \tan\beta \tan\beta' v_{\bar T}-2 \Lambda_T \chi_d v_S-2 \lambda \chi_u \tan\beta v_S), \nonumber\\
&&m_{H_u^0 {\bar T}^0}=\frac{1}{2} v_d (2 \sqrt2 \tan\beta T_{\chi_u}+\lambda (\Lambda_T \tan\beta v_{\bar T}-2 \chi_u v_S)-\tan\beta v_{\bar T} (g^2-8 \chi_u^2)), \nonumber\\
&&m_{S {\bar T}^0}=\Lambda_T^2 v_S v_{\bar T}-\frac{1}{2} \tan\beta' v_{\bar T} (\sqrt2 T_{\Lambda_T}+2 \kappa \Lambda_T v_S)-\frac{1}{2} v_d^2 (\Lambda_T \chi_d+2 \tan\beta \lambda \chi_u), \nonumber\\
&&m_{T^0 {\bar T}^0}=\frac{1}{2} (-\sqrt2 T_{\Lambda_T} v_S-2 g^2 \tan\beta' v_{\bar T}^2-\kappa \Lambda_T v_S^2+\lambda \Lambda_T \tan\beta v_d^2+2 \tan\beta' \Lambda_T^2 v_{\bar T}^2),
\end{eqnarray}
where $g^2=g_{_1}^2+g_{_2}^2$, $N=\sqrt2 T_{\lambda}+\kappa \lambda v_S+2 \lambda \chi_d \tan\beta' v_{\bar T}+2 \lambda \chi_u v_{\bar T}$, $x=\Lambda_T \chi_u \tan\beta \tan\beta' v_{\bar T}$, $n=\lambda v_{\bar T} (\tan\beta' \chi_d+\chi_u)$.

\section{The corresponding Wilson coefficients}

The one loop Wilson coefficients for the process $\bar B\rightarrow X_s\gamma$ are written as
\begin{eqnarray}
&&C_{7,NP}^{(a)}(\mu_{EW})=\sum_{H^-_i,u_j}\frac{s_W^2}{2e^2V^*_{ts}V_{tb}} \Big\{ \frac{1}{2}C_{H^-_i\bar s u_j}^R C_{H^-_i b \bar u_j}^{L}[-I_3(x_{u_j},x_{H^-_i})+I_4(x_{u_j},x_{H^-_i})]+\nonumber\\
&&\qquad\qquad\qquad\quad\frac{m_{u_j}}{m_b}C_{H^-_i\bar s u_j}^L C_{H^-_i b \bar u_j}^{L}[-I_1(x_{u_j},x_{H^-_i})+I_3(x_{u_j},x_{H^-_i})]\Big\},
\end{eqnarray}
\begin{eqnarray}
&&C_{7,NP}^{(b)}(\mu_{EW})=\sum_{H^-_j,u_i}\frac{s_W^2}{3e^2V^*_{ts}V_{tb}} \Big\{ \frac{1}{2}C_{H^-_j\bar s u_i}^R C_{H^-_j b \bar u_i}^{L}[-I_1(x_{u_i},x_{H^-_j})+2I_3(x_{u_i},x_{H^-_j})\nonumber\\
&&\qquad\qquad\qquad\quad-I_4(x_{u_i},x_{H^-_j})]+\frac{m_{u_i}}{m_b}C_{H^-_j\bar s u_i}^L C_{H^-_j b \bar u_i}^{L}[I_1(x_{u_i},x_{H^-_j})-I_2(x_{u_i},x_{H^-_j})\nonumber\\
&&\qquad\qquad\qquad\quad-I_3(x_{u_i},x_{H^-_j})]\Big\},
\end{eqnarray}
\begin{eqnarray}
&&C_{7,NP}^{(c)}(\mu_{EW})=\sum_{U^+_i,\chi^-_j}\frac{s_W^2}{3e^2V^*_{ts}V_{tb}} \Big\{ \frac{1}{2}C_{U^+_i\bar s \chi^-_j}^R C_{U^+_i b \bar \chi^-_j}^{L}[-I_3(x_{\chi^-_j},x_{U^+_i})+I_4(x_{\chi^-_j},x_{U^+_i})]+\nonumber\\
&&\qquad\qquad\qquad\quad\frac{m_{\chi^-_j}}{m_b}C_{U^+_i\bar s \chi^-_j}^L C_{U^+_i b \bar \chi^-_j}^{L}[-I_1(x_{\chi^-_j},x_{U^+_i})+I_3(x_{\chi^-_j},x_{U^+_i})]\Big\},
\end{eqnarray}
\begin{eqnarray}
&&C_{7,NP}^{(d)}(\mu_{EW})=\sum_{U^+_j,\chi^-_i}\frac{s_W^2}{2e^2V^*_{ts}V_{tb}} \Big\{ \frac{1}{2}C_{U^+_i\bar s \chi^-_i}^R C_{U^+_j b \bar \chi^-_i}^{L}[-I_1(x_{\chi^-_i},x_{U^+_j})+2I_2(x_{\chi^-_i},x_{U^+_j})\nonumber\\
&&\qquad\qquad\qquad\quad-I_4(x_{\chi^-_i},x_{U^+_j})]+\frac{m_{\chi^-_i}}{m_b}C_{U^+_j\bar s \chi^-_i}^L C_{U^+_j b \bar \chi^-_i}^{L}[I_1(x_{\chi^-_i},x_{U^+_j})-I_2(x_{\chi^-_i},x_{U^+_j})\nonumber\\
&&\qquad\qquad\qquad\quad-I_3(x_{\chi^-_i},x_{U^+_j})]\Big\},
\end{eqnarray}
\begin{eqnarray}
&&C_{7,NP}^{(\xi)}(\mu_{EW})=C_{7,NP}^{\prime (\xi)}(\mu_{EW})(L\leftrightarrow R), (\xi=a,b,c,d),
\end{eqnarray}
where $C_{XYZ}^{L,R}$ denote the constant parts of the interaction vertices about particles $XYZ$, L and R represent the left and right-handed parts respectively.

Denoting $x_i=\frac{m_i^2}{m_W^2}$, the concrete expressions for $I_k(k=1,...,4)$ can be given as:
\begin{eqnarray}
&&I_1(x_1,x_2)=\frac{1+{\rm ln} x_2}{(x_2-x_1)}+\frac{x_1 {\rm ln} x_1-x_2 {\rm ln} x_2}{(x_2-x_1)^2},\nonumber\\
&&I_2(x_1,x_2)=-\frac{1+{\rm ln} x_1}{(x_2-x_1)}-\frac{x_1 {\rm ln} x_1-x_2 {\rm ln} x_2}{(x_2-x_1)^2},\nonumber\\
&&I_3(x_1,x_2)=\frac{1}{2}\Big[\frac{3+2{\rm ln} x_2}{(x_2-x_1)}-\frac{2x_2+4x_2{\rm ln} x_2}{(x_2-x_1)^2}-\frac{2x_1^2{\rm ln} x_1}{(x_2-x_1)^3}+\frac{2x_2^2{\rm ln} x_2}{(x_2-x_1)^3}\Big],\nonumber\\
&&I_4(x_1,x_2)=\frac{1}{4}\Big[\frac{11+6{\rm ln} x_2}{(x_2-x_1)}-\frac{15+18x_2{\rm ln} x_2}{(x_2-x_1)^2}+\frac{6x_2^2+18x_2^2{\rm ln} x_2}{(x_2-x_1)^3}+\nonumber\\
&&\qquad\qquad\quad\frac{6x_1^3{\rm ln} x_1-6x_2^3{\rm ln} x_2}{(x_2-x_1)^4}\Big].
\end{eqnarray}

Assuming $m_{\chi_i^\pm},\;m_{\chi_j^0}\gg m_{_W}$, the two loop Wilson coefficients for the process $\bar B\rightarrow X_s\gamma$ can be given by
\begin{eqnarray}
&&C_{7,NP}^{WW}(\mu_{EW})=\sum_{\chi^\pm_i,\chi^0_j}\frac{-1}{8\pi^2} \Big\{ (|C_{W^-\bar\chi^0_j \chi^+_i}^{L}|^2 +|C_{W^-\bar\chi^0_j \chi^+_i}^{R}|^2)\Big[ P(\frac{1}{8},\frac{1}{4},\frac{-1}{48},\frac{-1}{144},1,x_t)+\nonumber\\
&&\qquad\qquad\qquad\quad \frac{2}{3}P(\frac{-11}{12},\frac{-29}{72},\frac{-1}{12},\frac{1}{144},1,x_t)\Big]+(C_{W^-\bar\chi^0_j \chi^+_i}^{R} C_{W^-\bar\chi^0_j \chi^+_i}^{L*}+C_{W^-\bar\chi^0_j \chi^+_i}^{L}\nonumber\\
&&\qquad\qquad\qquad\quad C_{W^-\bar\chi^0_j \chi^+_i}^{R*})\Big[ P(\frac{1}{16},\frac{1}{4},\frac{-1}{16},\frac{1}{144},1,x_t)+\frac{2}{3}P(\frac{-11}{12},\frac{-29}{72},
\frac{-1}{12},\frac{1}{144},1,\nonumber\\
&&\qquad\qquad\qquad\quad x_t)\Big]+\frac{1}{8}(C_{W^-\bar\chi^0_j \chi^+_i}^{L} C_{W^-\bar\chi^0_j \chi^+_i}^{R*}-C_{W^-\bar\chi^0_j \chi^+_i}^{R} C_{W^-\bar\chi^0_j \chi^+_i}^{L*})\frac{\partial^2 \rho_{2,1}}{\partial x_1^2}(x_W,x_t)\Big\}\nonumber\\
&&\qquad\qquad\qquad\quad+C^{L,R}_{W^-\bar\chi^+_i \chi^{++}}\leftrightarrow C^{L,R}_{W^-\bar\chi^0_j \chi^+_i},
\end{eqnarray}
\begin{eqnarray}
&&C_{8,NP}^{WW}(\mu_{EW})=\sum_{\chi^\pm_i,\chi^0_j}\frac{-3}{8\pi^2} \Big\{ (|C_{W^-\bar\chi^0_j \chi^+_i}^{L}|^2 +|C_{W^-\bar\chi^0_j \chi^+_i}^{R}|^2)P(\frac{-11}{12},\frac{-29}{72},\frac{-1}{12},\frac{1}{144},1,x_t)+\nonumber\\
&&\qquad\qquad\qquad\quad(C_{W^-\bar\chi^0_j \chi^+_i}^{R} C_{W^-\bar\chi^0_j \chi^+_i}^{L*}+C_{W^-\bar\chi^0_j \chi^+_i}^{L} C_{W^-\bar\chi^0_j \chi^+_i}^{R*})P(\frac{-1}{12},\frac{-5}{24},\frac{1}{12},\frac{-1}{144},1,x_t)+\nonumber\\
&&\qquad\qquad\qquad\quad(C_{W^-\bar\chi^0_j \chi^+_i}^{R} C_{W^-\bar\chi^0_j \chi^+_i}^{L*}-C_{W^-\bar\chi^0_j \chi^+_i}^{L} C_{W^-\bar\chi^0_j \chi^+_i}^{R*})P(\frac{1}{16},\frac{7}{24},0,0,1,x_t)\Big\}\nonumber\\
&&\qquad\qquad\qquad\quad+C^{L,R}_{W^-\bar\chi^+_i \chi^{++}}\leftrightarrow C^{L,R}_{W^-\bar\chi^0_j \chi^+_i},
\end{eqnarray}
\begin{eqnarray}
&&C_{7,NP}^{WH}(\mu_{EW})=\sum_{\chi^\pm_i,\chi^0_j,H_k^{\pm}}\frac{C_{H_k^-\bar d u}^{L}m_W^2}{4\sqrt{2}\pi^2m_dm_fV_{ud}^*}\Big\{\Big[\Re\Big(C_{H_k^-\bar\chi^0_j \chi^+_i}^{L} C_{W^-\bar\chi^0_j \chi^+_i}^{L}+C_{H_k^-\bar\chi^0_j \chi^+_i}^{R} C_{W^-\bar\chi^0_j \chi^+_i}^{R}\Big)-\nonumber\\
&&\qquad\qquad\qquad\quad i\Im\Big(C_{H_k^-\bar\chi^0_j \chi^+_i}^{L} C_{W^-\bar\chi^0_j \chi^+_i}^{L}+C_{H_k^-\bar\chi^0_j \chi^+_i}^{R} C_{W^-\bar\chi^0_j \chi^+_i}^{R}\Big)\Big]\Big(\frac{21}{64}-\frac{5}{288}+\frac{5}{24}J(m_W^2,\nonumber\\
&&\qquad\qquad\qquad\quad M_{H_k^{\pm}}^2,m_t^2)\Big)+\Big[\Re\Big(C_{H_k^-\bar\chi^0_j \chi^+_i}^{L} C_{W^-\bar\chi^0_j \chi^+_i}^{R}+C_{H_k^-\bar\chi^0_j \chi^+_i}^{R} C_{W^-\bar\chi^0_j \chi^+_i}^{L}\Big)-\nonumber\\
&&\qquad\qquad\qquad\quad i\Im\Big(C_{H_k^-\bar\chi^0_j \chi^+_i}^{L} C_{W^-\bar\chi^0_j \chi^+_i}^{R}+C_{H_k^-\bar\chi^0_j \chi^+_i}^{R} C_{W^-\bar\chi^0_j \chi^+_i}^{L}\Big)\Big]\Big(\frac{-1}{144}-\frac{1}{24}J(m_W^2,\nonumber\\
&&\qquad\qquad\qquad\quad M_{H_k^{\pm}}^2,m_t^2)\Big)-\Big[\Re\Big(C_{H_k^-\bar\chi^0_j \chi^+_i}^{L} C_{W^-\bar\chi^0_j \chi^+_i}^{L}-C_{H_k^-\bar\chi^0_j \chi^+_i}^{R} C_{W^-\bar\chi^0_j \chi^+_i}^{R}\Big)-\nonumber\\
&&\qquad\qquad\qquad\quad i\Im\Big(C_{H_k^-\bar\chi^0_j \chi^+_i}^{L} C_{W^-\bar\chi^0_j \chi^+_i}^{L}-C_{H_k^-\bar\chi^0_j \chi^+_i}^{R} C_{W^-\bar\chi^0_j \chi^+_i}^{R}\Big)\Big]\Big(\frac{16}{144}+\frac{1}{6}J(m_W^2,\nonumber\\
&&\qquad\qquad\qquad\quad M_{H_k^{\pm}}^2,m_t^2)\Big)-\Big[\Re\Big(C_{H_k^-\bar\chi^0_j \chi^+_i}^{L} C_{W^-\bar\chi^0_j \chi^+_i}^{R}-C_{H_k^-\bar\chi^0_j \chi^+_i}^{R} C_{W^-\bar\chi^0_j \chi^+_i}^{L}\Big)-\nonumber\\
&&\qquad\qquad\qquad\quad i\Im\Big(C_{H_k^-\bar\chi^0_j \chi^+_i}^{L} C_{W^-\bar\chi^0_j \chi^+_i}^{R}-C_{H_k^-\bar\chi^0_j \chi^+_i}^{R} C_{W^-\bar\chi^0_j \chi^+_i}^{L}\Big)\Big]\Big(\frac{1}{72}+\frac{1}{12}J(m_W^2,\nonumber\\
&&\qquad\qquad\qquad\quad M_{H_k^{\pm}}^2,m_t^2)\Big)\Big\}+C^{L,R}_{W^-(H_k^-)\bar\chi^+_i \chi^{++}}\leftrightarrow C^{L,R}_{W^-(H_k^-)\bar\chi^0_j \chi^+_i},
\end{eqnarray}
\begin{eqnarray}
&&C_{8,NP}^{WH}(\mu_{EW})=\sum_{\chi^\pm_i,\chi^0_j,H_k^{\pm}}\frac{C_{H_k^-\bar d u}^{L}m_W^2}{4\sqrt{2}\pi^2m_dm_fV_{ud}^*}\Big\{\Big[\Re\Big(C_{H_k^-\bar\chi^0_j \chi^+_i}^{L} C_{W^-\bar\chi^0_j \chi^+_i}^{L}+C_{H_k^-\bar\chi^0_j \chi^+_i}^{R} C_{W^-\bar\chi^0_j \chi^+_i}^{R}\Big)-\nonumber\\
&&\qquad\qquad\qquad\quad i\Im\Big(C_{H_k^-\bar\chi^0_j \chi^+_i}^{L} C_{W^-\bar\chi^0_j \chi^+_i}^{L}+C_{H_k^-\bar\chi^0_j \chi^+_i}^{R} C_{W^-\bar\chi^0_j \chi^+_i}^{R}\Big)\Big]\Big(\frac{-1}{8\sqrt{2}}J(m_W^2,M_{H_k^{\pm}}^2,m_t^2)\Big)+\nonumber\\
&&\qquad\qquad\qquad\quad \Big[\Re\Big(C_{H_k^-\bar\chi^0_j \chi^+_i}^{L} C_{W^-\bar\chi^0_j \chi^+_i}^{R}+C_{H_k^-\bar\chi^0_j \chi^+_i}^{R} C_{W^-\bar\chi^0_j \chi^+_i}^{L}\Big)+\nonumber\\
&&\qquad\qquad\qquad\quad i\Im\Big(C_{H_k^-\bar\chi^0_j \chi^+_i}^{L} C_{W^-\bar\chi^0_j \chi^+_i}^{R}+C_{H_k^-\bar\chi^0_j \chi^+_i}^{R} C_{W^-\bar\chi^0_j \chi^+_i}^{L}\Big)\Big]\Big(\frac{-1}{8\sqrt{2}}J(m_W^2,M_{H_k^{\pm}}^2,m_t^2)\Big)+\nonumber\\
&&\qquad\qquad\qquad\quad\Big[\Re\Big(C_{H_k^-\bar\chi^0_j \chi^+_i}^{L} C_{W^-\bar\chi^0_j \chi^+_i}^{L}-C_{H_k^-\bar\chi^0_j \chi^+_i}^{R} C_{W^-\bar\chi^0_j \chi^+_i}^{R}\Big)-\nonumber\\
&&\qquad\qquad\qquad\quad i\Im\Big(C_{H_k^-\bar\chi^0_j \chi^+_i}^{L} C_{W^-\bar\chi^0_j \chi^+_i}^{L}-C_{H_k^-\bar\chi^0_j \chi^+_i}^{R} C_{W^-\bar\chi^0_j \chi^+_i}^{R}\Big)\Big]\Big(\frac{-1}{4\sqrt{2}}J(m_W^2,M_{H_k^{\pm}}^2,m_t^2)\Big)+\nonumber\\
&&\qquad\qquad\qquad\quad\Big[\Re\Big(C_{H_k^-\bar\chi^0_j \chi^+_i}^{L} C_{W^-\bar\chi^0_j \chi^+_i}^{R}+C_{H_k^-\bar\chi^0_j \chi^+_i}^{R} C_{W^-\bar\chi^0_j \chi^+_i}^{L}\Big)+\nonumber\\
&&\qquad\qquad\qquad\quad i\Im\Big(C_{H_k^-\bar\chi^0_j \chi^+_i}^{L} C_{W^-\bar\chi^0_j \chi^+_i}^{R}+C_{H_k^-\bar\chi^0_j \chi^+_i}^{R} C_{W^-\bar\chi^0_j \chi^+_i}^{L}\Big)\Big]\Big(\frac{1}{4\sqrt{2}}J(m_W^2,\nonumber\\
&&\qquad\qquad\qquad\quad M_{H_k^{\pm}}^2,m_t^2)\Big)\Big\}+C^{L,R}_{W^-(H_k^-)\bar\chi^+_i \chi^{++}}\leftrightarrow C^{L,R}_{W^-(H_k^-)\bar\chi^0_j \chi^+_i}.
\end{eqnarray}

The concrete expressions for $P$ and $J$ are given by:
\begin{eqnarray}
&&\rho_{i,j}(x_1,x_2)=\frac{x_1^i \ln^j x_1-x_2^i \ln^j x_2}{x_1-x_2},\nonumber\\
&&P(y_1,y_2,y_3,y_4,x_1,x_2)=y_1\frac{\partial \rho_{1,1}(x_1,x_2)}{\partial x_1}+y_2\frac{\partial^2 \rho_{2,1}(x_1,x_2)}{\partial x_1^2}+\nonumber\\
&&\qquad\qquad\qquad\qquad\qquad\quad y_3\frac{\partial^3 \rho_{3,1}(x_1,x_2)}{\partial x_1^3}+y_4\frac{\partial^4 \rho_{4,1}(x_1,x_2)}{\partial x_1^4},\nonumber\\
&&J(x_1,x_2,x_3)=\ln m_F^2-\frac{\rho_{2,1}(x_1,x_3)-\rho_{2,2}(x_2,x_3)}{x_1^2-x_2^2},
\end{eqnarray}
where $m_F$ runs all $m_{\chi_i^\pm}$, $m_{\chi_j^0}$.

The one loop Wilson coefficients for the process $B_s^0\rightarrow \mu^+\mu^-$ are written as
\begin{eqnarray}
&&C_{_{S,NP}}^{(1)}(\mu_{_{\rm EW}})=\sum_{_{\tilde U_i,\chi^-_j,\chi^-_k,S=h_l,A_l
}}\frac{C_{\mu^-S\mu^+}^L+C_{\mu^-S\mu^+}^R}{2(m_b^2-m_S^2)}\Big[C_{\tilde U_i\bar s \chi^-_j}^R C_{\bar \chi^-_j S \chi^-_k}^L C_{\bar \chi^-_k s \tilde U_i}^R G_2(x_{\tilde U_i},x_{\chi^\pm_j},x_{\chi^\pm_k})\nonumber\\
&&\qquad\qquad\qquad+M_{\chi^\pm_j}M_{\chi^\pm_k} C_{\tilde U_i\bar s \chi^-_j}^R C_{\bar \chi^-_j S \chi^-_k}^R C_{\bar \chi^-_k s \tilde U_i}^R G_1(x_{\tilde U_i},x_{\chi^\pm_j},x_{\chi^\pm_k})\Big]\nonumber\\
&&\qquad\qquad\qquad+\sum_{_{H^-_i,u_j,u_k,S=h_l,A_l}}\frac{C_{\mu^-S\mu^+}^L+C_{\mu^-S\mu^+}^R}{2(m_b^2-m_S^2)}\Big[C_{H^-_i\bar s u_j}^R C_{\bar u_j S u_k}^L C_{\bar u_k b H^-_i}^R G_2(x_{\tilde H^\pm_i},x_{u_j},x_{u_k})\nonumber\\
&&\qquad\qquad\qquad+m_{u_j}m_{u_k}C_{H^-_i\bar s u_j}^R C_{\bar u_j S u_k}^R C_{\bar u_k b H^-_i}^R G_1(x_{\tilde H^\pm_i},x_{u_j},x_{u_k})\Big],\nonumber\\
&&C_{_{P,NP}}^{(1)}(\mu_{_{\rm EW}})=\sum_{_{\tilde U_i,\chi^-_j,\chi^-_k,S=h_l,A_l
}}\frac{-C_{\mu^-S\mu^+}^L+C_{\mu^-S\mu^+}^R}{2(m_b^2-m_S^2)}\Big[C_{\tilde U_i\bar s \chi^-_j}^R C_{\bar \chi^-_j S \chi^-_k}^L C_{\bar \chi^-_k s \tilde U_i}^R G_2(x_{\tilde U_i},x_{\chi^\pm_j},x_{\chi^\pm_k})\nonumber\\
&&\qquad\qquad\qquad+M_{\chi^\pm_j}M_{\chi^\pm_k} C_{\tilde U_i\bar s \chi^-_j}^R C_{\bar \chi^-_j S \chi^-_k}^R C_{\bar \chi^-_k s \tilde U_i}^R G_1(x_{\tilde U_i},x_{\chi^\pm_j},x_{\chi^\pm_k})\Big]\nonumber\\
&&\qquad\qquad\qquad+\sum_{_{H^-_i,u_j,u_k,S=h_l,A_l}}\frac{-C_{\mu^-S\mu^+}^L+C_{\mu^-S\mu^+}^R}{2(m_b^2-m_S^2)}\Big[C_{H^-_i\bar s u_j}^R C_{\bar u_j S u_k}^L C_{\bar u_k b H^-_i}^R G_2(x_{\tilde H^\pm_i},x_{u_j},x_{u_k})\nonumber\\
&&\qquad\qquad\qquad+m_{u_j}m_{u_k}C_{H^-_i\bar s u_j}^R C_{\bar u_j S u_k}^R C_{\bar u_k b H^-_i}^R G_1(x_{\tilde H^\pm_i},x_{u_j},x_{u_k})\Big],
\end{eqnarray}
\begin{eqnarray}
&&C_{_{S,NP}}^{(2)}(\mu_{_{\rm EW}})=\sum_{u_i,H^\pm_j,H^\pm_k,S=h_l,A_l}\frac{1}{2(m_b^2-m_S^2)}m_{u_i}C_{\bar s u_i H^\pm_j}^RC_{\bar u_i b H^\pm_k}^RC_{S H^\pm_j H^\pm_k}G_1(x_{u_i},x_{H^\pm_j},x_{H^\pm_k})\nonumber\\
&&\qquad\qquad\qquad (C_{\mu^-S\mu^+}^L+C_{\mu^-S\mu^+}^R)\nonumber\\
&&\qquad\qquad\qquad+\sum_{\chi^\pm_i,\tilde U_j,\tilde U_k,S=h_l,A_l}\frac{1}{2(m_b^2-m_S^2)}m_{\chi^\pm_i}C_{\bar s \chi^\pm_i \tilde U_j}^RC_{\bar \chi^\pm_i b \tilde U_k}^RC_{S \tilde U_j \tilde U_k}G_1(x_{\chi^\pm_i},x_{\tilde U_j},x_{\tilde U_k})\nonumber\\
&&\qquad\qquad\qquad (C_{\mu^-S\mu^+}^L+C_{\mu^-S\mu^+}^R),\nonumber\\
&&C_{_{p,NP}}^{(2)}(\mu_{_{\rm EW}})=\sum_{u_i,H^\pm_j,H^\pm_k,S=h_l,A_l}\frac{1}{2(m_b^2-m_S^2)}m_{u_i}C_{\bar s u_i H^\pm_j}^RC_{\bar u_i b H^\pm_k}^RC_{S H^\pm_j H^\pm_k}G_1(x_{u_i},x_{H^\pm_j},x_{H^\pm_k})\nonumber\\
&&\qquad\qquad\qquad (-C_{\mu^-S\mu^+}^L+C_{\mu^-S\mu^+}^R)\nonumber\\
&&\qquad\qquad\qquad+\sum_{\chi^\pm_i,\tilde U_j,\tilde U_k,S=h_l,A_l}\frac{1}{2(m_b^2-m_S^2)}m_{\chi^\pm_i}C_{\bar s \chi^\pm_i \tilde U_j}^RC_{\bar \chi^\pm_i b \tilde U_k}^RC_{S \tilde U_j \tilde U_k}G_1(x_{\chi^\pm_i},x_{\tilde U_j},x_{\tilde U_k})\nonumber\\
&&\qquad\qquad\qquad (-C_{\mu^-S\mu^+}^L+C_{\mu^-S\mu^+}^R),
\end{eqnarray}
\begin{eqnarray}
&&C_{_{S,NP}}^{(3)}(\mu_{_{\rm EW}})=\sum_{u_i,H^\pm_k,S=h_l,A_l}\frac{-C_{W^\pm S H^\pm_k}}{2(m_b^2-m_S^2)}\Big[C_{\bar s W^\pm u_i}^LC_{\bar u_i H^\pm_k b}^RG_2(x_{u_i},1,x_{H^\pm_k})-2m_bm_{u_i}C_{\bar s W^\pm u_i}^L\nonumber\\
&&\qquad\qquad\qquad C_{\bar u_i H^\pm_k b}^LG_1(x_{u_i},1,x_{H^\pm_k})\Big](C_{\mu^-S\mu^+}^L+C_{\mu^-S\mu^+}^R),\nonumber\\
&&C_{_{P,NP}}^{(3)}(\mu_{_{\rm EW}})=\sum_{u_i,H^\pm_k,S=h_l,A_l}\frac{-C_{W^\pm S H^\pm_k}}{2(m_b^2-m_S^2)}\Big[C_{\bar s W^\pm u_i}^LC_{\bar u_i H^\pm_k b}^RG_2(x_{u_i},1,x_{H^\pm_k})-2m_bm_{u_i}C_{\bar s W^\pm u_i}^L\nonumber\\
&&\qquad\qquad\qquad C_{\bar u_i H^\pm_k b}^LG_1(x_{u_i},1,x_{H^\pm_k})\Big](-C_{\mu^-S\mu^+}^L+C_{\mu^-S\mu^+}^R),
\end{eqnarray}
\begin{eqnarray}
&&C_{_{S,NP}}^{(4)}(\mu_{_{\rm EW}})=\sum_{u_i,H^\pm_j,S=h_l,A_l}\frac{-C_{W^\pm S H^\pm_j}}{2(m_b^2-m_S^2)}C_{\bar s H^\pm_j u_i}^RC_{\bar u_i W^\pm b}^RG_2(x_{u_i},x_{H^\pm_j},1)(C_{\mu^-S\mu^+}^L+C_{\mu^-S\mu^+}^R),\nonumber\\
&&C_{_{S,NP}}^{(4)}(\mu_{_{\rm EW}})=\sum_{u_i,H^\pm_j,S=h_l,A_l}\frac{-C_{W^\pm S H^\pm_j}}{2(m_b^2-m_S^2)}C_{\bar s H^\pm_j u_i}^RC_{\bar u_i W^\pm b}^RG_2(x_{u_i},x_{H^\pm_j},1)(-C_{\mu^-S\mu^+}^L+C_{\mu^-S\mu^+}^R),\nonumber\\
\end{eqnarray}
\begin{eqnarray}
&&C_{_{9,NP}}^{(5)}(\mu_{_{\rm EW}})=\sum_{_{\tilde U_i,\chi^-_j,\chi^-_k,V
}}\frac{C_{\mu^-V\mu^+}^L+C_{\mu^-V\mu^+}^R}{-2(m_b^2-m_V^2)}\Big[-\frac{1}{2}C_{\tilde U_i\bar s \chi^-_j}^R C_{\bar \chi^-_j V \chi^-_k}^R C_{\bar \chi^-_k s \tilde U_i}^L G_2(x_{\tilde U_i},x_{\chi^\pm_j},x_{\chi^\pm_k})\nonumber\\
&&\qquad\qquad\qquad+M_{\chi^\pm_j}M_{\chi^\pm_k} C_{\tilde U_i\bar s \chi^-_j}^R C_{\bar \chi^-_j V \chi^-_k}^L C_{\bar \chi^-_k s \tilde U_i}^L G_1(x_{\tilde U_i},x_{\chi^\pm_j},x_{\chi^\pm_k})\Big]\nonumber\\
&&\qquad\qquad\qquad+\sum_{_{\tilde H^\pm_i,u_j,u_k,V
}}\frac{C_{\mu^-V\mu^+}^L+C_{\mu^-V\mu^+}^R}{-2(m_b^2-m_V^2)}\Big[-\frac{1}{2}C_{H^\pm_i\bar s u_j}^R C_{\bar u_j V u_k}^R C_{u_k s H^\pm_i}^L G_2(x_{H^\pm_i},x_{u_j},x_{u_k})\nonumber\\
&&\qquad\qquad\qquad+m_{u_j}m_{u_k} C_{H^\pm_i\bar s u_j}^R C_{\bar u_j V u_k}^L C_{\bar u_k s H^\pm_i}^L G_1(x_{H^\pm_i},x_{u_j},x_{u_k})\Big],\nonumber\\
&&C_{_{10,NP}}^{(5)}(\mu_{_{\rm EW}})=\sum_{_{\tilde U_i,\chi^-_j,\chi^-_k,V
}}\frac{-C_{\mu^-V\mu^+}^L+C_{\mu^-V\mu^+}^R}{-2(m_b^2-m_V^2)}\Big[-\frac{1}{2}C_{\tilde U_i\bar s \chi^-_j}^R C_{\bar \chi^-_j V \chi^-_k}^R C_{\bar \chi^-_k s \tilde U_i}^L G_2(x_{\tilde U_i},x_{\chi^\pm_j},x_{\chi^\pm_k})\nonumber\\
&&\qquad\qquad\qquad+M_{\chi^\pm_j}M_{\chi^\pm_k} C_{\tilde U_i\bar s \chi^-_j}^R C_{\bar \chi^-_j V \chi^-_k}^L C_{\bar \chi^-_k s \tilde U_i}^L G_1(x_{\tilde U_i},x_{\chi^\pm_j},x_{\chi^\pm_k})\Big]\nonumber\\
&&\qquad\qquad\qquad+\sum_{_{\tilde H^\pm_i,u_j,u_k,V
}}\frac{-C_{\mu^-V\mu^+}^L+C_{\mu^-V\mu^+}^R}{-2(m_b^2-m_V^2)}\Big[-\frac{1}{2}C_{H^\pm_i\bar s u_j}^R C_{\bar u_j V u_k}^R C_{u_k s H^\pm_i}^L G_2(x_{H^\pm_i},x_{u_j},x_{u_k})\nonumber\\
&&\qquad\qquad\qquad+m_{u_j}m_{u_k} C_{H^\pm_i\bar s u_j}^R C_{\bar u_j V u_k}^L C_{\bar u_k s H^\pm_i}^L G_1(x_{H^\pm_i},x_{u_j},x_{u_k})\Big],
\end{eqnarray}
\begin{eqnarray}
&&C_{_{9,NP}}^{(6)}(\mu_{_{\rm EW}})=\sum_{_{u_i,H^\pm_j,H^\pm_k,V
}}\frac{C_{\mu^-V\mu^+}^L+C_{\mu^-V\mu^+}^R}{4(m_b^2-m_V^2)}C_{\bar s u_i H^\pm_j}^RC_{\bar u_i b H^\pm_k}^LC_{V H^\pm_j H^\pm_k}G_2(x_{u_i},x_{H^\pm_j},x_{H^\pm_k})\nonumber\\
&&\qquad\qquad\qquad+\sum_{_{\chi^\pm_i,\tilde U_j,\tilde U_k,V
}}\frac{C_{\mu^-V\mu^+}^L+C_{\mu^-V\mu^+}^R}{4(m_b^2-m_V^2)}C_{\bar s \chi^\pm_i \tilde U_j}^RC_{\bar \chi^\pm_i b \tilde U_k}^LC_{V \tilde U_j \tilde U_k}G_2(x_{\chi^\pm_i},x_{\tilde U_j},x_{\tilde U_k}),\nonumber\\
&&C_{_{10,NP}}^{(6)}(\mu_{_{\rm EW}})=\sum_{_{u_i,H^\pm_j,H^\pm_k,V
}}\frac{-C_{\mu^-V\mu^+}^L+C_{\mu^-V\mu^+}^R}{4(m_b^2-m_V^2)}C_{\bar s u_i H^\pm_j}^RC_{\bar u_i b H^\pm_k}^LC_{V H^\pm_j H^\pm_k}G_2(x_{u_i},x_{H^\pm_j},x_{H^\pm_k})\nonumber\\
&&\qquad\qquad\qquad+\sum_{_{\chi^\pm_i,\tilde U_j,\tilde U_k,V
}}\frac{-C_{\mu^-V\mu^+}^L+C_{\mu^-V\mu^+}^R}{4(m_b^2-m_V^2)}C_{\bar s \chi^\pm_i \tilde U_j}^RC_{\bar \chi^\pm_i b \tilde U_k}^LC_{V \tilde U_j \tilde U_k}G_2(x_{\chi^\pm_i},x_{\tilde U_j},x_{\tilde U_k}),\nonumber\\
&&C_{_{S,NP}}^{(6)}(\mu_{_{\rm EW}})=\sum_{_{u_i,H^\pm_j,H^\pm_k,V
}}\frac{C_{\mu^-V\mu^+}^L+C_{\mu^-V\mu^+}^R}{-2(m_b^2-m_V^2)}m_bm_{u_i}C_{\bar s u_i H^\pm_j}^RC_{\bar u_i b H^\pm_k}^RC_{V H^\pm_j H^\pm_k}G_1(x_{u_i},x_{H^\pm_j},x_{H^\pm_k})\nonumber\\
&&\qquad\qquad\qquad+\sum_{_{\chi^\pm_i,\tilde U_j,\tilde U_k,V
}}\frac{C_{\mu^-V\mu^+}^L+C_{\mu^-V\mu^+}^R}{-2(m_b^2-m_V^2)}m_bm_{\chi^\pm_i}C_{\bar s \chi^\pm_i \tilde U_j}^RC_{\bar \chi^\pm_i b \tilde U_k}^RC_{V \tilde U_j \tilde U_k}G_1(x_{\chi^\pm_i},x_{\tilde U_j},x_{\tilde U_k}),\nonumber\\
&&C_{_{P,NP}}^{(6)}(\mu_{_{\rm EW}})=\sum_{_{u_i,H^\pm_j,H^\pm_k,V
}}\frac{C_{\mu^-V\mu^+}^L-C_{\mu^-V\mu^+}^R}{-2(m_b^2-m_V^2)}m_bm_{u_i}C_{\bar s u_i H^\pm_j}^RC_{\bar u_i b H^\pm_k}^RC_{V H^\pm_j H^\pm_k}G_1(x_{u_i},x_{H^\pm_j},x_{H^\pm_k})\nonumber\\
&&\qquad\qquad\qquad+\sum_{_{\chi^\pm_i,\tilde U_j,\tilde U_k,V
}}\frac{C_{\mu^-V\mu^+}^L-C_{\mu^-V\mu^+}^R}{-2(m_b^2-m_V^2)}m_bm_{\chi^\pm_i}C_{\bar s \chi^\pm_i \tilde U_j}^RC_{\bar \chi^\pm_i b \tilde U_k}^RC_{V \tilde U_j \tilde U_k}G_1(x_{\chi^\pm_i},x_{\tilde U_j},x_{\tilde U_k}),\nonumber\\
\end{eqnarray}
\begin{eqnarray}
&&C_{_{9,NP}}^{(7)}(\mu_{_{\rm EW}})=\sum_{_{u_i,H^\pm_k,V
}}\frac{C_{\mu^-V\mu^+}^L+C_{\mu^-V\mu^+}^R}{2(m_b^2-m_V^2)}m_{u_i}C_{\bar s u_i W^\pm}^LC_{\bar u_i b H^\pm_k}^LC_{V W^\pm H^\pm_k}G_1(x_{u_i},x_W,x_{H^\pm_k}),\nonumber\\
&&C_{_{10,NP}}^{(7)}(\mu_{_{\rm EW}})=\sum_{_{u_i,H^\pm_k,V
}}\frac{-C_{\mu^-V\mu^+}^L+C_{\mu^-V\mu^+}^R}{2(m_b^2-m_V^2)}m_{u_i}C_{\bar s u_i W^\pm}^LC_{\bar u_i b H^\pm_k}^LC_{V W^\pm H^\pm_k}G_1(x_{u_i},x_W,x_{H^\pm_k}),\nonumber\\
\end{eqnarray}
\begin{eqnarray}
&&C_{_{9,NP}}^{(8)}(\mu_{_{\rm EW}})=\sum_{_{u_i,H^\pm_j,V
}}\frac{C_{\mu^-V\mu^+}^L+C_{\mu^-V\mu^+}^R}{2(m_b^2-m_V^2)}m_{u_i}C_{\bar s u_i H^\pm_j}^RC_{\bar u_i b W^\pm}^LC_{V W^\pm H^\pm_k}G_1(x_{u_i},x_{H^\pm_j},x_W),\nonumber\\
&&C_{_{10,NP}}^{(8)}(\mu_{_{\rm EW}})=\sum_{_{u_i,H^\pm_j,V
}}\frac{-C_{\mu^-V\mu^+}^L+C_{\mu^-V\mu^+}^R}{2(m_b^2-m_V^2)}m_{u_i}C_{\bar s u_i H^\pm_j}^RC_{\bar u_i b W^\pm}^LC_{V W^\pm H^\pm_k}G_1(x_{u_i},x_{H^\pm_j},x_W),\nonumber\\
\end{eqnarray}
\begin{eqnarray}
&&C_{_{9,NP}}^{(9)}(\mu_{_{\rm EW}})=\sum_{_{\tilde U_i,\chi^\pm_j,\chi^\pm_k,\tilde\nu_l
}}-\frac{1}{8}C_{\bar s \tilde U_i \chi^\pm_j}^RC_{\bar \chi^\pm_j \mu^+ \tilde\nu_l}^L(C_{\bar\mu^-\tilde\nu_l\chi^\pm_k}^LC_{\bar\chi^\pm_k \tilde U_i b}^R+C_{\bar\mu^-\tilde\nu_l\chi^\pm_k}^RC_{\bar\chi^\pm_k \tilde U_i b}^L)\nonumber\\
&&\qquad\qquad\qquad G_4(x_{\tilde U_i},x_{\chi^\pm_j},x_{\chi^\pm_k},x_{\tilde \nu_l}),\nonumber\\
&&C_{_{10,NP}}^{(9)}(\mu_{_{\rm EW}})=\sum_{_{\tilde U_i,\chi^\pm_j,\chi^\pm_k,\tilde\nu_l
}}-\frac{1}{8}C_{\bar s \tilde U_i \chi^\pm_j}^RC_{\bar \chi^\pm_j \mu^+ \tilde\nu_l}^L(C_{\bar\mu^-\tilde\nu_l\chi^\pm_k}^LC_{\bar\chi^\pm_k \tilde U_i b}^R-C_{\bar\mu^-\tilde\nu_l\chi^\pm_k}^RC_{\bar\chi^\pm_k \tilde U_i b}^L)\nonumber\\
&&\qquad\qquad\qquad G_4(x_{\tilde U_i},x_{\chi^\pm_j},x_{\chi^\pm_k},x_{\tilde \nu_l}),\nonumber\\
&&C_{_{S,NP}}^{(9)}(\mu_{_{\rm EW}})=\sum_{_{\tilde U_i,\chi^\pm_j,\chi^\pm_k,\tilde\nu_l
}}-\frac{1}{2}M_{\chi^\pm_j}M_{\chi^\pm_k}C_{\bar s \tilde U_i \chi^\pm_j}^RC_{\bar \chi^\pm_j \mu^+ \tilde\nu_l}^R(C_{\bar\mu^-\tilde\nu_l\chi^\pm_k}^LC_{\bar\chi^\pm_k \tilde U_i b}^L+C_{\bar\mu^-\tilde\nu_l\chi^\pm_k}^RC_{\bar\chi^\pm_k \tilde U_i b}^R)\nonumber\\
&&\qquad\qquad\qquad G_3(x_{\tilde U_i},x_{\chi^\pm_j},x_{\chi^\pm_k},x_{\tilde \nu_l}),\nonumber\\
&&C_{_{P,NP}}^{(9)}(\mu_{_{\rm EW}})=\sum_{_{\tilde U_i,\chi^\pm_j,\chi^\pm_k,\tilde\nu_l
}}-\frac{1}{2}M_{\chi^\pm_j}M_{\chi^\pm_k}C_{\bar s \tilde U_i \chi^\pm_j}^RC_{\bar \chi^\pm_j \mu^+ \tilde\nu_l}^R(-C_{\bar\mu^-\tilde\nu_l\chi^\pm_k}^LC_{\bar\chi^\pm_k \tilde U_i b}^L+C_{\bar\mu^-\tilde\nu_l\chi^\pm_k}^RC_{\bar\chi^\pm_k \tilde U_i b}^R)\nonumber\\
&&\qquad\qquad\qquad G_3(x_{\tilde U_i},x_{\chi^\pm_j},x_{\chi^\pm_k},x_{\tilde \nu_l}),
\end{eqnarray}
\begin{eqnarray}
&&C_{_{9,NP}}^{(10)}(\mu_{_{\rm EW}})=\sum_{_{u_i,H^\pm_j,H^\pm_k,\tilde\nu_l}}\frac{1}{8}C_{\bar s u_i H^\pm_j}^RC_{\bar u_i b H^\pm_k}^L(C_{\bar\mu^-H^\pm_k\nu_l}^LC_{\bar\nu_l H^\pm_j \mu^+}^R+C_{\bar\mu^-H^\pm_k\nu_l}^RC_{\bar\nu_l H^\pm_j \mu^+}^L)\nonumber\\
&&\qquad\qquad\qquad G_4(x_{u_i},x_{H^\pm_j},x_{H^\pm_k},x_{\nu_l}),\nonumber\\
&&C_{_{10,NP}}^{(10)}(\mu_{_{\rm EW}})=\sum_{_{u_i,H^\pm_j,H^\pm_k,\tilde\nu_l}}\frac{1}{8}C_{\bar s u_i H^\pm_j}^RC_{\bar u_i b H^\pm_k}^L(C_{\bar\mu^-H^\pm_k\nu_l}^LC_{\bar\nu_l H^\pm_j \mu^+}^R-C_{\bar\mu^-H^\pm_k\nu_l}^RC_{\bar\nu_l H^\pm_j \mu^+}^L)\nonumber\\
&&\qquad\qquad\qquad G_4(x_{u_i},x_{H^\pm_j},x_{H^\pm_k},x_{\nu_l}),
\end{eqnarray}
\begin{eqnarray}
&&C_{_{S,NP}}^{(11)}(\mu_{_{\rm EW}})=-\sum_{_{u_i,H^\pm_j,\nu_l}}\frac{1}{2}C_{\bar s u_i H^\pm_j}^RC_{\bar u_i b W^\pm}^RC_{\bar\mu^-W^\pm\nu_l}^RC_{\bar\nu_l H^\pm_j \mu^+}^L G_4(x_{u_i},x_{H^\pm_j},x_W,x_{\nu_l}),\nonumber\\
&&C_{_{P,NP}}^{(11)}(\mu_{_{\rm EW}})=-\sum_{_{u_i,H^\pm_j,\nu_l}}\frac{1}{2}C_{\bar s u_i H^\pm_j}^RC_{\bar u_i b W^\pm}^RC_{\bar\mu^-W^\pm\nu_l}^RC_{\bar\nu_l H^\pm_j \mu^+}^L G_4(x_{u_i},x_{H^\pm_j},x_W,x_{\nu_l}),
\end{eqnarray}
where $V$ denotes photon $\gamma$, $Z$ boson and $C_{\bar tVt},\;C_{\bar\chi_j^0V\chi_j^0},\;C_{\bar\chi_i^+V\chi_i^+}$ denote the vector parts of the corresponding interaction vertex.

The two loop Wilson coefficients for the process $B_s^0\rightarrow \mu^+\mu^-$ are written as
\begin{eqnarray}
&&C_{_{9,NP}}^{WW}(\mu_{_{\rm EW}})=\sum_{_{\chi^+_i,\chi^0_j,V}}\frac{(C_{\mu^-V\mu^+}^R+C_{\mu^-V\mu^+}^L)g_{s}^2}{128\pi^4 s_W^2}\frac{m_b^2}{m_b^2-m_V^2} \Big\{ \Big[(|C_{W^-\bar\chi^0_j \chi^+_i}^{L}|^2 +|C_{W^-\bar\chi^0_j \chi^+_i}^{R}|^2) \nonumber\\
&&\qquad\qquad\qquad\quad P(\frac{1}{8},\frac{1}{4},\frac{-1}{48},\frac{-1}{144},1,x_t)+(|C_{W^-\bar\chi^0_j \chi^+_i}^{L}|^2 -|C_{W^-\bar\chi^0_j \chi^+_i}^{R}|^2)\frac{\partial^1 \rho_{1,1}}{\partial x_1}(x_W,x_t)+\nonumber\\
&&\qquad\qquad\qquad\quad(C_{W^-\bar\chi^0_j \chi^+_i}^{R}C_{W^-\bar\chi^0_j \chi^+_i}^{L*}+C_{W^-\bar\chi^0_j \chi^+_i}^{L}C_{W^-\bar\chi^0_j \chi^+_i}^{R*}) P(\frac{1}{16},\frac{1}{4},\frac{-1}{16},\frac{1}{144},1,x_t)+\nonumber\\
&&\qquad\qquad\qquad\quad \frac{1}{8}(C_{W^-\bar\chi^0_j \chi^+_i}^{L} C_{W^-\bar\chi^0_j \chi^+_i}^{R*}-C_{W^-\bar\chi^0_j \chi^+_i}^{R} C_{W^-\bar\chi^0_j \chi^+_i}^{L*})\frac{\partial^2 \rho_{2,1}}{\partial x_1^2}(x_W,x_t)\Big]C_{VW^-W^-}+\nonumber\\
&&\qquad\qquad\qquad\quad\Big[(|C_{W^-\bar\chi^0_j \chi^+_i}^{L}|^2 +|C_{W^-\bar\chi^0_j \chi^+_i}^{R}|^2)P(\frac{1}{144},\frac{-1}{12},\frac{-29}{72},\frac{-11}{12},1,x_t)+(C_{W^-\bar\chi^0_j \chi^+_i}^{R}\nonumber\\
&&\qquad\qquad\qquad\quad C_{W^-\bar\chi^0_j \chi^+_i}^{L*}+C_{W^-\bar\chi^0_j \chi^+_i}^{L}C_{W^-\bar\chi^0_j \chi^+_i}^{R*}) P(\frac{-1}{144},\frac{1}{12},\frac{-5}{24},\frac{-1}{12},1,x_t)\Big]C_{\bar tVt}+\nonumber\\
&&\qquad\qquad\qquad\quad \Big[(|C_{W^-\bar\chi^0_j \chi^+_i}^{L}|^2 +|C_{W^-\bar\chi^0_j \chi^+_i}^{R}|^2)P(\frac{3}{16},\frac{3}{16},0,0,1,x_t)+\frac{1}{16}(C_{W^-\bar\chi^0_j \chi^+_i}^{L}\nonumber\\
&&\qquad\qquad\qquad\quad C_{W^-\bar\chi^0_j \chi^+_i}^{R*}-C_{W^-\bar\chi^0_j \chi^+_i}^{R} C_{W^-\bar\chi^0_j \chi^+_i}^{L*})\frac{\partial^2 \rho_{2,1}}{\partial x_1^2}(x_W,x_t)\Big](C_{\bar\chi_j^0V\chi_j^0}+C_{\bar\chi_i^+V\chi_i^+})\Big\}\nonumber\\
&&\qquad\qquad\qquad\quad+C^{L,R}_{W^-\bar\chi^+_i \chi^{++}}\leftrightarrow C^{L,R}_{W^-\bar\chi^0_j \chi^+_i},
\end{eqnarray}
\begin{eqnarray}
&&C_{_{10,NP}}^{WW}(\mu_{_{\rm EW}})=\sum_{_{\chi^+_i,\chi^0_j,V}}\frac{(C_{\mu^-V\mu^+}^R-C_{\mu^-V\mu^+}^L)g_{s}^2}{128\pi^4 s_W^2}\frac{m_b^2}{m_b^2-m_V^2} \Big\{ \Big[(|C_{W^-\bar\chi^0_j \chi^+_i}^{L}|^2 +|C_{W^-\bar\chi^0_j \chi^+_i}^{R}|^2) \nonumber\\
&&\qquad\qquad\qquad\quad P(\frac{1}{8},\frac{1}{4},\frac{-1}{48},\frac{-1}{144},1,x_t)+(|C_{W^-\bar\chi^0_j \chi^+_i}^{L}|^2 -|C_{W^-\bar\chi^0_j \chi^+_i}^{R}|^2)\frac{\partial^1 \rho_{1,1}}{\partial x_1}(x_W,x_t)+\nonumber\\
&&\qquad\qquad\qquad\quad(C_{W^-\bar\chi^0_j \chi^+_i}^{R}C_{W^-\bar\chi^0_j \chi^+_i}^{L*}+C_{W^-\bar\chi^0_j \chi^+_i}^{L}C_{W^-\bar\chi^0_j \chi^+_i}^{R*}) P(\frac{1}{16},\frac{1}{4},\frac{-1}{16},\frac{1}{144},1,x_t)+\nonumber\\
&&\qquad\qquad\qquad\quad \frac{1}{8}(C_{W^-\bar\chi^0_j \chi^+_i}^{L} C_{W^-\bar\chi^0_j \chi^+_i}^{R*}-C_{W^-\bar\chi^0_j \chi^+_i}^{R} C_{W^-\bar\chi^0_j \chi^+_i}^{L*})\frac{\partial^2 \rho_{2,1}}{\partial x_1^2}(x_W,x_t)\Big]C_{VW^-W^-}+\nonumber\\
&&\qquad\qquad\qquad\quad\Big[(|C_{W^-\bar\chi^0_j \chi^+_i}^{L}|^2 +|C_{W^-\bar\chi^0_j \chi^+_i}^{R}|^2)P(\frac{1}{144},\frac{-1}{12},\frac{-29}{72},\frac{-11}{12},1,x_t)+(C_{W^-\bar\chi^0_j \chi^+_i}^{R}\nonumber\\
&&\qquad\qquad\qquad\quad C_{W^-\bar\chi^0_j \chi^+_i}^{L*}+C_{W^-\bar\chi^0_j \chi^+_i}^{L}C_{W^-\bar\chi^0_j \chi^+_i}^{R*}) P(\frac{-1}{144},\frac{1}{12},\frac{-5}{24},\frac{-1}{12},1,x_t)\Big]C_{\bar tVt}+\nonumber\\
&&\qquad\qquad\qquad\quad \Big[(|C_{W^-\bar\chi^0_j \chi^+_i}^{L}|^2 +|C_{W^-\bar\chi^0_j \chi^+_i}^{R}|^2)P(\frac{3}{16},\frac{3}{16},0,0,1,x_t)+\frac{1}{16}(C_{W^-\bar\chi^0_j \chi^+_i}^{L}\nonumber\\
&&\qquad\qquad\qquad\quad C_{W^-\bar\chi^0_j \chi^+_i}^{R*}-C_{W^-\bar\chi^0_j \chi^+_i}^{R} C_{W^-\bar\chi^0_j \chi^+_i}^{L*})\frac{\partial^2 \rho_{2,1}}{\partial x_1^2}(x_W,x_t)\Big](C_{\bar\chi_j^0V\chi_j^0}+C_{\bar\chi_i^+V\chi_i^+})\Big\}\nonumber\\
&&\qquad\qquad\qquad\quad+C^{L,R}_{W^-\bar\chi^+_i \chi^{++}}\leftrightarrow C^{L,R}_{W^-\bar\chi^0_j \chi^+_i}.
\end{eqnarray}

The concrete expressions for $G_k(k=1,...,4)$ are written as
\begin{eqnarray}
&&G_1(x_1,x_2,x_3)=\frac{-1}{m_W^2}\Big[\frac{x_1{\rm ln} x_1}{(x_2-x_1)(x_3-x_1)}+\frac{x_2{\rm ln} x_2}{(x_1-x_2)(x_3-x_2)}+\frac{x_3{\rm ln} x_3}{(x_1-x_3)(x_2-x_3)}\Big],\nonumber\\
&&G_2(x_1,x_2,x_3)=-\frac{x_1^2{\rm ln} x_1}{(x_2-x_1)(x_3-x_1)}-\frac{x_2^2{\rm ln} x_2}{(x_1-x_2)(x_3-x_2)}-\frac{x_3^2{\rm ln} x_3}{(x_1-x_3)(x_2-x_3)},\nonumber\\
&&G_3(x_1,x_2,x_3,x_4)=\frac{1}{m_W^4}\Big[\frac{x_1{\rm ln} x_1}{(x_2-x_1)(x_3-x_1)(x_4-x_1)}+\frac{x_2{\rm ln} x_2}{(x_1-x_2)(x_3-x_2)(x_4-x_2)}+\nonumber\\
&&\qquad\qquad\qquad\qquad\frac{x_3{\rm ln} x_3}{(x_1-x_3)(x_2-x_3)(x_4-x_3)}\frac{x_4{\rm ln} x_4}{(x_1-x_4)(x_2-x_4)(x_3-x_4)}\Big],\nonumber\\
&&G_4(x_1,x_2,x_3,x_4)=\frac{1}{m_W^2}\Big[\frac{x_1^2{\rm ln} x_1}{(x_2-x_1)(x_3-x_1)(x_4-x_1)}+\frac{x_2^2{\rm ln} x_2}{(x_1-x_2)(x_3-x_2)(x_4-x_2)}+\nonumber\\
&&\qquad\qquad\qquad\qquad\frac{x_3^2{\rm ln} x_3}{(x_1-x_3)(x_2-x_3)(x_4-x_3)}\frac{x_4^2{\rm ln} x_4}{(x_1-x_4)(x_2-x_4)(x_3-x_4)}\Big].\nonumber\\
\end{eqnarray}

\end{document}